\documentclass[a4paper,10pt]{article}
\pdfoutput=1
\usepackage{graphicx,rotating,hyperref,slashed,xcolor,amsmath,amssymb,amsfonts,colortbl,cite,subfigure,float,mathrsfs,mathalpha}
\makeatletter 
\usepackage{graphics}
\usepackage{ulem} 
\usepackage{comment} 
\graphicspath{{plots/}}
 
\hypersetup{colorlinks,bookmarksopen,bookmarksnumbered,
linkcolor=blus,pdfstartview=FitH,urlcolor=rossos,citecolor=verde}
\numberwithin{equation}{section}

\hypersetup{%
    ,urlcolor=rossos
    ,citecolor=rossos
    ,linkcolor=rossos
    }

\AtBeginDocument{
  \hypersetup{
    urlcolor=verdes,
    citecolor=verdes,
    linkcolor=verdes,
  }%
}

\allowdisplaybreaks

\def\lsim{\mathrel{\rlap{\lower3pt\hbox{\hskip0pt$\sim$}}
   \raise1pt\hbox{$<$}}}         
\def\gsim{\mathrel{\rlap{\lower4pt\hbox{\hskip1pt$\sim$}}
   \raise1pt\hbox{$>$}}}         

 \newcommand{\sfootnote}[1]{} 
\definecolor{bluc}{cmyk}{1,1,0,0.1}
\definecolor{rossoCP3}{cmyk}{0,.88,.77,.40}
\definecolor{rosso}{cmyk}{0,1,1,0.4}
\definecolor{rossos}{cmyk}{0,1,1,0.55}
\definecolor{rossoc}{cmyk}{0,1,1,0.2}
\definecolor{verdes}{cmyk}{0.92,0,0.59,0.4}

\hypersetup{colorlinks, bookmarksopen, bookmarksnumbered,
citecolor=verdes, linkcolor=bluc, pdfstartview=FitH, urlcolor=rossos}

\newcommand{\mio}[1]{}

\definecolor{Gray}{gray}{0.95}

\newcommand{\pb}{\,{\rm pb}}

\usepackage{multicol}
\usepackage{color}
\definecolor{rosso}{cmyk}{0,1,1,0.4}
\definecolor{rossos}{cmyk}{0,1,1,0.55}
\definecolor{rossoc}{cmyk}{0,1,1,0.2}
\definecolor{blu}{cmyk}{1,1,0,0.3}
\definecolor{blus}{cmyk}{1,1,0,0.6}
\definecolor{bluc}{cmyk}{1,1,0,0.1}
\definecolor{verde}{cmyk}{0.92,0,0.59,0.25}
\definecolor{verdec}{cmyk}{0.92,0,0.59,0.15}
\definecolor{verdes}{cmyk}{0.92,0,0.59,0.4}

\def\circa#1{\,\raise.3ex\hbox{$#1$\kern-.75em\lower1ex\hbox{$\sim$}}\,}

\newcommand{\beq}{\begin{equation}}
\newcommand{\eeq}{\end{equation}}

\newcommand{\bea}{\begin{eqnarray}}
\newcommand{\eea}{\end{eqnarray}}
\newcommand{\be}{\begin{equation}}
\newcommand{\ee}{\end{equation}}
\newfam\rsfsfam
\def\mathscr#1{{\fam\rsfsfam\relax#1}}

\def\circa#1{\,\raise.3ex\hbox{$#1$\kern-.75em\lower1ex\hbox{$\sim$}}\,}
\makeatletter

\def\hhref#1{\href{http://arxiv.org/abs/#1}{arXiv:#1}} 

\newcommand{\doi}[1]{\href{http://dx.doi.org/#1}{[doi]}}

\def\hhref#1{\href{http://arxiv.org/abs/#1}{arXiv:#1}} 
 
\def\art{\@ifnextchar[{\eart}{\oart}}
\def\eart[#1]#2#3#4#5#6{{\rm #2}, {\em #3 \bf #4} {\rm (#6) #5} ({\em #1})}

\def\article{\@ifnextchar[{\earticle}{\oarticle}}
\def\oarticle#1#2#3#4#5#6{{\rm #1}, {\em ``#6''}, {\rm #2 #3 (#5) #4}}
\def\earticle[#1]#2#3#4#5#6#7{{\rm #2}, {\em ``#7''}, {\rm #3 #4 (#6) #5}  [\hhref{#1}]}
\def\hepart[#1]#2{{\rm #2, \em#1}}
\def\heparticle[#1]#2#3{#2, {\em ``#3''} [\hhref{#1}]}

\newcounter{alphaequation}[equation]
\def\thealphaequation{\theequation\hbox to
0.6em{\hfil\alph{alphaequation}\hfil}}
\def\eqnsystem#1{
\def\@eqnnum{{\rm (\thealphaequation)}}
\def\@@eqncr{\let\@tempa\relax \ifcase\@eqcnt \def\@tempa{& & &} \or
  \def\@tempa{& &}\or \def\@tempa{&}\fi\@tempa
  \if@eqnsw\@eqnnum\refstepcounter{alphaequation}\fi
\global\@eqnswtrue\global\@eqcnt=0\cr}
\refstepcounter{equation} \let\@currentlabel\theequation \def\@tempb{#1}
\ifx\@tempb\empty\else\label{#1}\fi
\refstepcounter{alphaequation}
\let\@currentlabel\thealphaequation
\global\@eqnswtrue\global\@eqcnt=0 \tabskip\@centering\let\\=\@eqncr
$$\halign to \displaywidth\bgroup \@eqnsel\hskip\@centering
$\displaystyle\tabskip\z@{##}$&\global\@eqcnt\@ne
\hskip2\arraycolsep\hfil${##}$\hfil& \global\@eqcnt\tw@\hskip2\arraycolsep
$\displaystyle\tabskip\z@{##}$\hfil
\tabskip\@centering&\llap{##}\tabskip\z@\cr}
\def\endeqnsystem{\@@eqncr\egroup$$\global\@ignoretrue} \makeatother

\definecolor{fiorentina}{rgb}{.5,0,.5}

\def \bm#1{\mbox{\boldmath$#1$\unboldmath}}

\def\be{\begin{equation}}
\def\ee{\end{equation}}
\def\bea{\begin{eqnarray}}
\def\eea{\end{eqnarray}}

\def\f{\frac}

\def\d{{\rm d}}
\def\Mpl{M_{_{\mathrm{Pl}}}}
\def\mpcinv{{\rm Mpc}^{-1}}

\def\cAk{{\cal A}_k}
\def\vx{{\bm{x}}}
\def\vk{{\bm{k}}}
\def\vq{{\bm{q}}}

\def\pt{\mathcal{P}_{_{\mathrm{T}}}}

\def\pb{\mathcal{P}_{_{\mathrm{B}}}}
\def\pe{\mathcal{P}_{_{\mathrm{E}}}}

\def\ee{\eta_{\rm e}}

\def\ogw{\Omega_{_{\mathrm{GW}}}}

\def\f{\frac}

\begin{document}

\begin{center} 
{\selectfont \sffamily \bfseries \Large 
Chiral gravitational waves from multi-phase magnetogenesis
}
\vskip 0.5cm
{H.~V.~Ragavendra$^{a,b\!}$, Gianmassimo Tasinato$^{c,d\!}$, and L.~Sriramkumar$^{e\!}$}
\vskip 0.5cm
{\small 
\textsl{$^a$Dipartimento di Fisica e Astronomia “Galileo Galilei”, Universit\`{a} degli Studi di Padova, \\ Via Marzolo 8, I-35131 Padova, Italy}
\\
\textsl{$^b$Istituto Nazionale di Fisica Nucleare (INFN), Sezione di Padova, Via Marzolo 8, I-35131 Padova, Italy}
\\
\textsl{$^c$Physics Department, Swansea University, SA2 8PP, UK}
\\
\textsl{$^d$Dipartimento di Fisica e Astronomia, Universit\`a di Bologna,\\
 INFN, Sezione di Bologna,  viale B. Pichat 6/2, 40127 Bologna,   Italy}\\
\textsl{$^e$Centre for Strings, Gravitation and Cosmology,
Department of Physics,\\
Indian Institute of Technology Madras, Chennai~600036, India}}
\end{center}
\vskip 0.5cm
\begin{abstract}
Cosmological vector fields are central to many early-Universe phenomena, including inflationary dynamics, primordial magnetogenesis, and dark-matter scenarios. However, constructing models able to generate cosmological  magnetic fields while avoiding strong coupling, backreaction, and cosmic microwave background constraints remains challenging. We study a novel mechanism in which brief non--slow-roll phases during inflation amplify  primordial magnetic fields at small scales, while maintaining theoretical consistency and observational viability.  We  incorporate parity-violating interactions in the vector sector and demonstrate, for the first time in  a non--slow-roll framework, that chirality can significantly boost magnetic-field amplitudes and imprint distinctive polarization-dependent spectral features. We complement detailed numerical computations with an analytical treatment yielding compact expressions for chiral vector mode functions that  reproduce the main spectral properties. We then develop a systematic formalism to evaluate the stochastic gravitational-wave background naturally induced at second order by these amplified fields, identifying both an intensity component and a circularly polarized contribution with characteristic frequency profiles. We discuss detection prospects with future multiband gravitational-wave observatories, showing that chiral signatures could provide a distinctive observational probe. Our results introduce new avenues for enhancing primordial magnetic fields and their associated gravitational-wave signals, opening promising possibilities for their future detection and interpretation,
both with cosmological and gravitational wave probes.
\end{abstract}

\section{Introduction}
\label{sec_intro}

Cosmological vector fields play a central role in a wide range of early-Universe phenomena, from models of inflation (see e.g.~\cite{Maleknejad:2012fw} for a review) to scenarios of primordial magnetogenesis (see~\cite{Turner:1987bw,Ratra:1991bn,Finelli:2000sh,Maroto:2000zu,
Maroto:2001ki,Matarrese:2004kq,Martin:2007ue,Demozzi:2009fu,Kanno:2009ei,Bamba:2006ga,Bamba:2008hr,Barnaby:2012tk,Ferreira:2013sqa,Ferreira:2014hma} for seminal works) and dark-matter models (see e.g.~\cite{Graham:2015rva}). In particular, primordial magnetogenesis provides an appealing mechanism to account for the large-scale magnetic fields observed in galaxies and clusters. However, concrete realizations  face significant theoretical and observational challenges, including strong-coupling regimes, backreaction effects, and  bounds from the cosmic microwave background (CMB); see e.g.~\cite{Durrer:2013pga,Subramanian:2015lua} for reviews. Developing scenarios that overcome these obstacles while still producing observable magnetic fields remains an important open problem.

In this work we explore a novel framework designed to amplify small-scale primordial magnetic fields while simultaneously generating chirality and satisfying both observational constraints (notably those from the CMB) and theoretical consistency requirements such as backreaction control. 
Our mechanism exploits a brief phase of departure of the inflaton from slow-roll evolution, which can influence the effective couplings between scalar and vector sectors. 
Such non--slow-roll evolutions are known to enhance the spectrum of curvature perturbations---potentially leading to primordial black-hole formation (see e.g.~\cite{Ozsoy:2023ryl} for a review)---and can likewise be used to boost inflationary magnetic fields. 
For the first time within a non--slow-roll setting, we consistently incorporate parity-violating effects in the vector sector and show that they can provide an additional and significant enhancement of the magnetic-field amplitude beyond that produced by the background dynamics alone, as well as adding new features
to the magnetic field profile. 

After introducing the set-up in Section~\ref{sec:model_numerics}, we perform in Section~\ref{sec_scen1} a detailed numerical analysis of the resulting magnetic-field spectrum. We demonstrate that the spectrum grows from large to small scales, extending and refining previous studies~\cite{Tripathy:2021sfb,Tripathy:2022iev,Atkins:2025pvg} of its  growth properties. We also show that this behavior allows the model to evade both backreaction constraints and current CMB bounds. We then include parity-violating interactions and find that chirality significantly modifies the spectral shape, generating additional polarization-dependent peaks whose features distinguish the two helicity states. In Section~\ref{sec_scen2} we complement the numerical study with an analytical approach based on~\cite{Tasinato:2020vdk,Tasinato:2023ukp}, deriving compact formulas for the chiral vector mode functions. Despite their simplicity, these expressions successfully reproduce the main characteristics of the full numerical spectra obtained in Section~\ref{sec_scen1}.

In Section~\ref{sec_vigw} we develop a systematic framework to compute the stochastic gravitational-wave background (SGWB) induced at second order by the enhanced magnetic fields on small scales. We show that the resulting energy-density spectrum can be decomposed into a standard contribution $\ogw$, governed by the total GW intensity, and a parity-violating component $\ogw^{V}$ associated with circular polarization. Both contributions exhibit distinctive spectral shapes that can lead to observable signatures. We discuss prospects for detection and characterization with future multiband gravitational-wave observatories, emphasizing that chirality could be inferred from characteristic spectral features. Overall, our scenario provides new mechanisms for enhancing the GW signal and endowing it with properties that may facilitate its detection and interpretation.


\section{Our set-up}
\label{sec:model_numerics}
Our work aims to develop both a numerical and analytical understanding of the properties of chiral magnetic fields generated during inflationary scenarios  featuring
a departure from slow-roll evolution. As we shall learn, a transient non-slow-roll epoch can lead to a significant rapid  amplification of magnetic fields at small scales. This enhancement has important consequences for the generation of a background of chiral gravitational waves after  inflation ends, which are sourced by the amplified magnetic field spectrum.

\smallskip

We work in a Friedmann--Lemaître--Robertson--Walker (FLRW) background 
(with a mostly-plus signature) and we  denote conformal time by $\tau$.  
The dynamics of the electromagnetic field during inflation is governed by the action
\cite{Sorbo:2011rzq,Caprini:2014mja}
\begin{equation}
S[A^\mu] = -\frac{1}{4} \int \d\tau\, \d^3\vx \, \sqrt{-g}\, J^2\,
\bigg[
F_{\mu\nu} F^{\mu\nu}
- \frac{\gamma}{2}\, F_{\mu\nu}\, \tilde F^{\mu\nu}
\bigg] \, ,
\label{eq:action-2}
\end{equation}
where the electromagnetic field strength tensor is defined in terms of the vector potential as
$F_{\mu\nu} = \partial_\mu A_\nu - \partial_\nu A_\mu$.
Its dual is given by
\begin{equation}
\tilde F^{\mu\nu} \equiv \frac{1}{\sqrt{-g}}\, \epsilon^{\mu\nu\alpha\beta} F_{\alpha\beta} \, ,
\end{equation}
with $\epsilon^{\mu\nu\alpha\beta}$ the totally antisymmetric Levi--Civita tensor.
The function $J$, if dependent
on coordinates, breaks the conformal invariance of the electromagnetic action allowing
for vector-field amplification, while the parameter $\gamma$ controls the strength of parity violation in the vector sector. 

During inflation, both $J$ and $\gamma$ are typically modelled as functions of the inflaton or of a spectator field, leading to direct couplings between scalar degrees of freedom and the electromagnetic field strength~\cite{Caprini:2014mja,Ng:2014lyb,Tasinato:2014fia,Subramanian:2009fu,
Sharma:2018kgs,Chowdhury:2018mhj,Okano:2020uyr,Brandenburg:2021bfx,Tripathy:2021sfb,Tripathy:2022iev,Tripathy:2024ngu,Papanikolaou:2024cwr}.
In scenarios that include brief non-slow-roll phases -- often motivated by primordial black hole production (see, e.g.~\cite{Ragavendra:2023ret} for a review), or
by punctuated inflation models~\cite{Bhaumik:2019tvl,Ragavendra:2020sop} -- the scalar field velocities may vary rapidly over short time intervals. 
This behaviour induces {\it sharp time variations} in both $J$ and $\gamma$ during the inflationary process.
Besides magnetogenesis, such scenarios are useful for building models of longitudinal
vector dark matter \cite{Marriott-Best:2025sez,LaRosa:2025woi}. 
In our set-up, without committing to explicit model building, we treat these quantities as explicit functions of conformal time $\tau$. We analyze two representative scenarios that capture their evolution in the presence of brief non-slow-roll epochs, 
during which the profiles of $J$ and $\gamma$ can change significantly. 
The class of models we wish to capture—such as those featuring violations of slow-roll in scenarios of primordial black hole formation—share the characteristic that the decaying mode, which would otherwise be strongly suppressed on superhorizon scales, becomes significantly excited.
A central result of our analysis is that, independently of the specific form or parametrization of the functions $J$ and $\gamma$, the resulting spectra exhibit universal behaviour. In particular, during a brief transition phase associated with sharp variations of these functions, we find that the slope of the spectrum follows a model-independent prediction. This universality manifests itself, for instance, in the existence of a bound on the maximal spectral slope, as well as in common qualitative features of the spectral profile.
It is precisely this robustness and insensitivity to model details that render our results both general and compelling, and ensure their applicability across a wide class of scenarios.

\smallskip

We  work in the Coulomb gauge setting $A_0=\partial_iA^i=0$.
We introduce the quantity $\cAk = J\,A_k$, where $A_k$ is the mode function 
in Fourier space, so that its corresponding equation of motion is~\cite{Sorbo:2011rzq,Tripathy:2021sfb,Tripathy:2022iev}
\begin{eqnarray}
\cAk^{\lambda\,\prime\prime} + \left(k^2 +2 \lambda\gamma\,k\f{J'}{J} - \f{J''}{J}\right)\cAk^\lambda &=& 0\,,
\label{eq:cAk-helical}
\end{eqnarray}
where the index of polarization is $\lambda=[L,R]=\pm1$,
and a prime indicates derivative along conformal time. 
While the term $J''/J \propto \tau^{-2}$ introduces a demarcation of sub-Hubble and 
super-Hubble behavior to the mode function,
the term $2 \lambda\gamma\,k J'/J$ induces chiral imbalance, enhancing the mode function
of one polarization, while suppressing the other.
The exact time, scale and chiral dependence of the effects of these terms is associated with  
 the behavior of $J(\tau)$.
The magnetic power spectrum $\pb^\lambda(k)$ for a given polarization
index $\lambda$ is then~\cite{Martin:2007ue,Subramanian:2009fu,Caprini:2014mja}
\begin{equation}
\pb^\lambda(k) = \f{J^2}{a^4}\f{k^5}{2\pi^2}\vert A_k^\lambda\vert^2
= \f{k^5}{2\pi^2a^4}\vert \cAk^\lambda\vert^2\,,
\end{equation}
which is evaluated at the end of inflation, when $J$ is set to reach unity, restoring conformal invariance.
For reference, the parity-preserving case ($\gamma=0$) with monotonic  evolution of 
$J = (a^2/a_{\rm e}^2)$  throughout inflation (with $a_\mathrm{e}$ 
being the scale factor at the end of inflation) leads to equal and scale-invariant 
magnetic field spectra for both helicities as~\cite{Tripathy:2021sfb,Tripathy:2024ngu}
\begin{equation}
\pb^L(k) = \pb^R(k) = \f{9H^4}{8\pi^2}\,.
\end{equation}
For $\gamma=0$
we shall denote as $\pb^0$ the corresponding total power spectrum  $\pb^{\gamma\to0}(k)=\pb^L(k)+\pb^R(k)=9H^4/(4\pi^2)$,  and use it as as reference to contrast against the features induced due to 
deviating phase of $J$ and/or parity-violation ($\gamma\neq0$), which is the
topic we are mostly interested in. 

\smallskip
In fact, we now pass
to discuss two complementary scenarios of inflationary magnetogenesis, differing
in the behaviour of $\gamma$ as a function of time. 
In  Section~\ref{sec_scen1} we investigate a first scenario, in which we extend the analysis of Ref.~\cite{Atkins:2025pvg} by accounting for the finite duration of the non-slow-roll epoch, and by including parity-violating effects through the parameter $\gamma$ appearing in Eq.~\eqref{eq:action-2}. We  make the hypothesis
that $\gamma$ is a constant parameter
turned on during the entire inflationary epoch. Our study of this case relies on a numerical analysis of the coupled mode equations, and it reveals new non-trivial features in the  profile of the resulting magnetic field spectrum. 
The second scenario, discussed in Section~\ref{sec_scen2}, further generalizes \cite{Atkins:2025pvg} by incorporating parity violation in a setup that remains analytically tractable. In this case, we assume that $\gamma$
is switched on {\it only}\/  during the brief non-slow-roll phase. We then derive a fully analytic expression for the gauge-field mode functions, as well as for the chiral magnetic field power spectrum. The interesting  implications of both scenarios for the generation of chiral induced gravitational waves are developed in Section~\ref{sec_vigw}.

\section{First scenario,  and numerical analysis}
\label{sec_scen1}

In the first scenario, we model the behavior of the non-conformal coupling function $J$ 
as often done in models of inflation with a brief phase of ultra slow roll (USR) sandwiched 
between slow-roll evolutions -- see e.g. the review~\cite{Ragavendra:2023ret} and 
references therein. We model the time evolution of $J$ as involving three phases, such 
as $J(\tau) \sim \tau^{-2},\,\tau^{-n_2},\,\tau^{-2}$.
The phases of $J \propto \tau^{-2}$ shall lead to scale-invariant $\pb(k)$,
while the intermediate phase shall lead to deviations from scale-invariance. 
We may expect such a behavior of $J$ when it is modelled in terms of the inflaton 
$\phi$ or the associated kinetic term $X=-\partial_\mu\phi\partial^\mu\phi/2$ in USR models or models that
have brief departure from slow-roll evolution (see, for related efforts~\cite{Tasinato:2014fia,Tripathy:2022iev,Tripathy:2024ngu}).

The exact form of $J(\tau)$ can be arrived at by matching conditions at transitions 
between phases, while ensuring continuity of the function: 
\begin{eqnarray}
J(\tau) &=&
\begin{cases}
\left(\f{\tau_{\rm e}}{\tau_2}\right)^{2}\left(\f{\tau_1}{\tau_2}\right)^{-n_2}
\left(\f{\tau}{\tau_1}\right)^{-2} & {\rm for}~\tau<\tau_1\,,  \\
\left(\f{\tau_{\rm e}}{\tau_2}\right)^2\left(\f{\tau}{\tau_2}\right)^{-n_2} & {\rm for}~\tau_1<\tau<\tau_2\,, \\
\left(\f{\tau}{\tau_{\rm e}}\right)^{-2} & {\rm for}~\tau>\tau_2\,,
\end{cases}
\end{eqnarray}
where $\tau_1$ and $\tau_2$ are the times  demarcating the phase junctions.
We approximate the transition between the three phases as instantaneous. 
We  ensure that $J(\tau)$ is continuous across phases and set $J(\tau_{\rm e})=1$
at the end of inflation $\tau_{\rm e}$\,.
The setup is very similar to the construction presented in~\cite{Atkins:2025pvg}.
The difference in the  scenario of this section is that here the intermediate phase is not instantaneous
but of finite duration, and the index describing the phase $n_2$ is of finite value. This fact has interesting
consequences for phenomenology.  
For numerical purposes, we describe  the transitions through a step function modelled 
using a hyperbolic tangent function in terms of e-folds
$N \equiv \ln(a/a(\tau_{\rm e}))$\,.

We solve the electromagnetic mode function numerically in terms of e-folds $N$
for the three-phase model of $J$ and compute the corresponding $\pb(k)$ at the 
end of inflation.
For carrying out numerics, the inflationary background dynamics can effectively 
be set as de Sitter expansion, parameterized by a constant  Hubble parameter $H$.
The scale factor is then $a(\tau)=-1/(H \tau)$, with $\tau<0$
during inflation. 
\subsection{Non-helical case: steepest $n_{\rm B}$}

First, we analyze the three-phase model in the parity-preserving case ($\gamma=0$), focusing on the steepest possible growth of $\pb(k)$. This question is also of interest in comparison with the scalar case. For curvature perturbations, the growing mode during non--slow-roll evolution is well described by a power-law spectrum with spectral index $n \simeq 4$, which represents the maximal slope attainable within this class of scenarios~\cite{Byrnes:2018txb,Ozsoy:2019lyy}. 

In contrast, the amplification of vector fluctuations exhibits a more intricate behavior. Although the magnetic power spectrum also grows steeply, with an effective spectral index initially close to $n \simeq 4$, its evolution is not accurately described by a pure power law. Instead, the slope can exceed that of the scalar case, reaching a maximal value of $n \simeq 4.75$, as derived analytically in~\cite{Atkins:2025pvg} under the assumption of an infinitesimally brief departure from slow roll. Here we show that even larger values of $n$ can be obtained when the non--slow-roll phase has a finite duration.

Specifically, we numerically analyze a setup featuring a short intermediate non--slow-roll phase of $J$ with a large power-law index $\tau^{n_2}$, following the construction of~\cite{Atkins:2025pvg}. We characterize this intermediate stage by large negative values $n_2 \in [-20,-26]$ and a brief duration $\Delta N_2 \leq 0.5$. For these parameters we compute $\pb(k)$ and the corresponding spectral index
\(
n_{\rm B} \equiv \frac{\mathrm{d}\ln \pb(k)}{\mathrm{d}\ln k}
\)
numerically, scanning the parameter space to determine the largest attainable value of $n_{\rm B}$.

\begin{figure}[t!]
\centering
\includegraphics[width=0.47\linewidth]{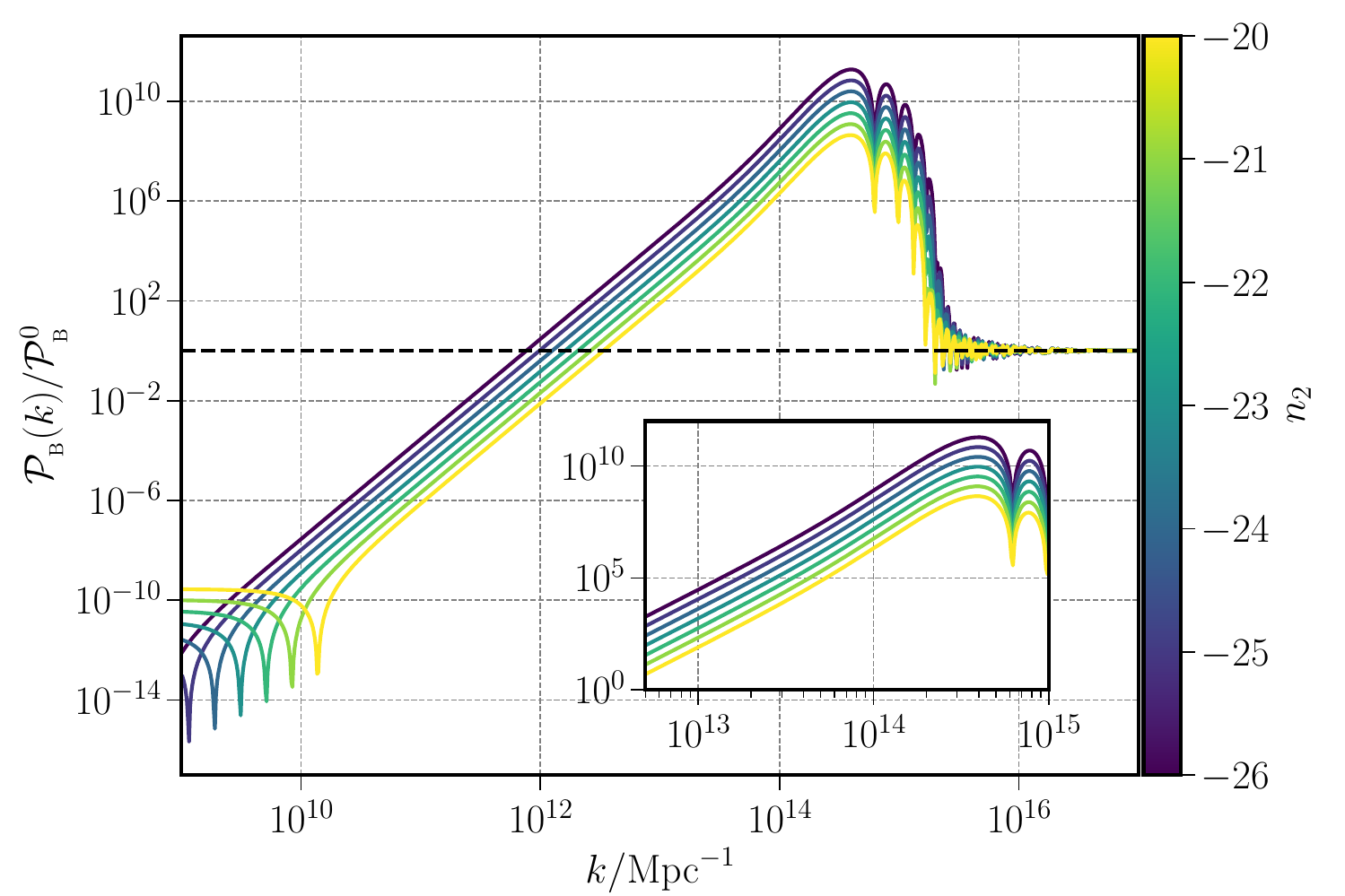}
\includegraphics[width=0.47\linewidth]{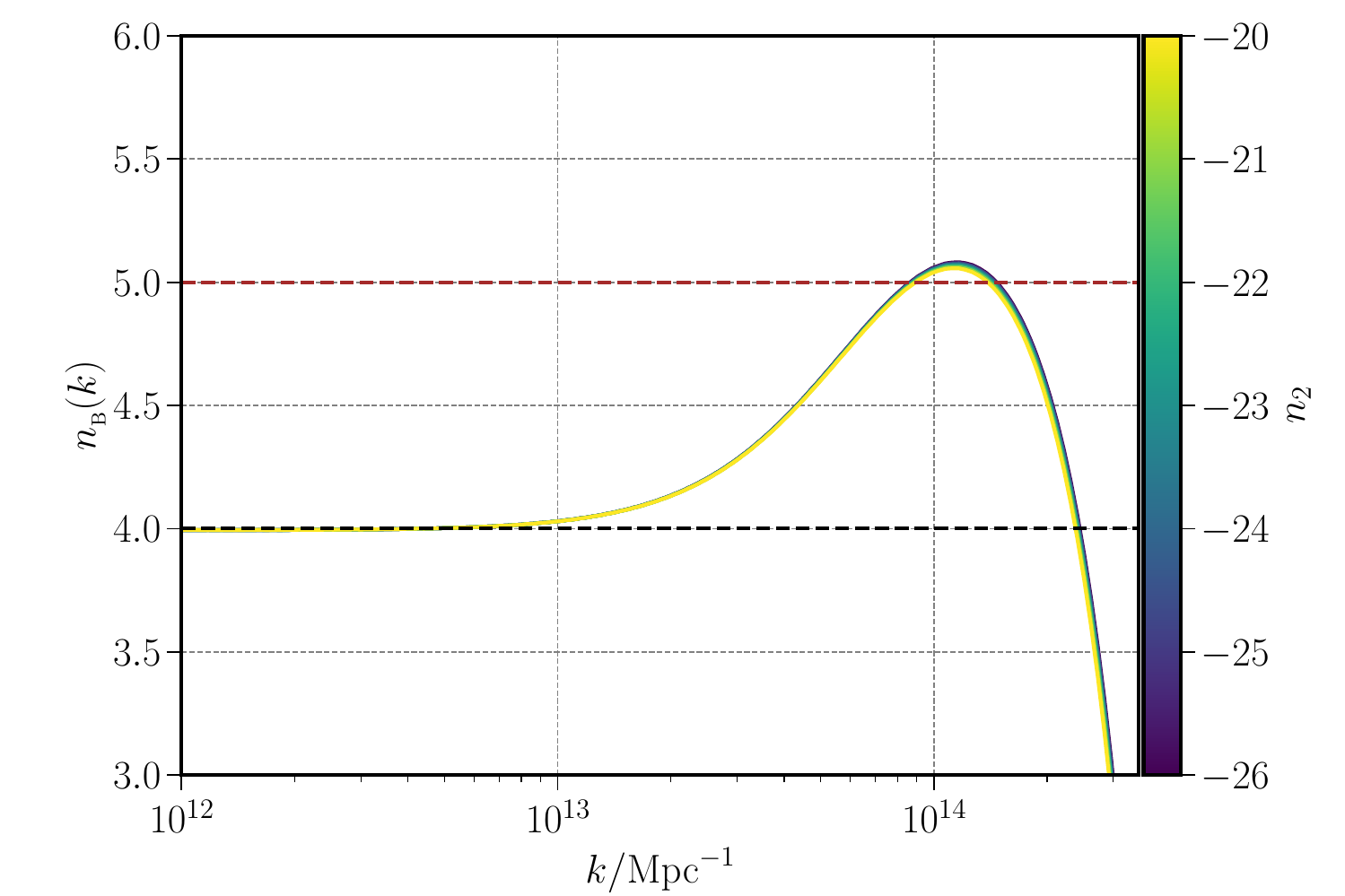}
\includegraphics[width=0.47\linewidth]{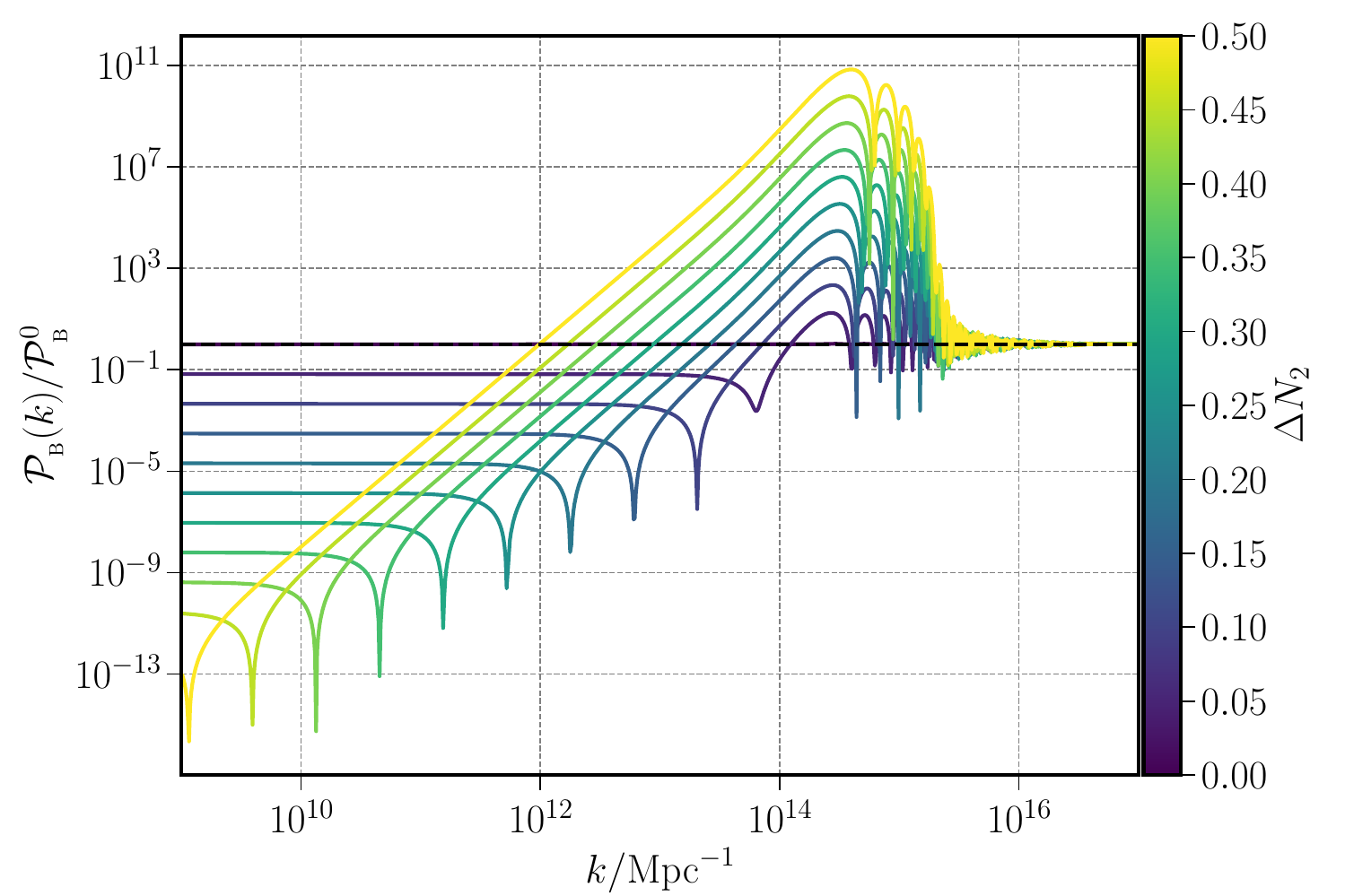}
\includegraphics[width=0.47\linewidth]{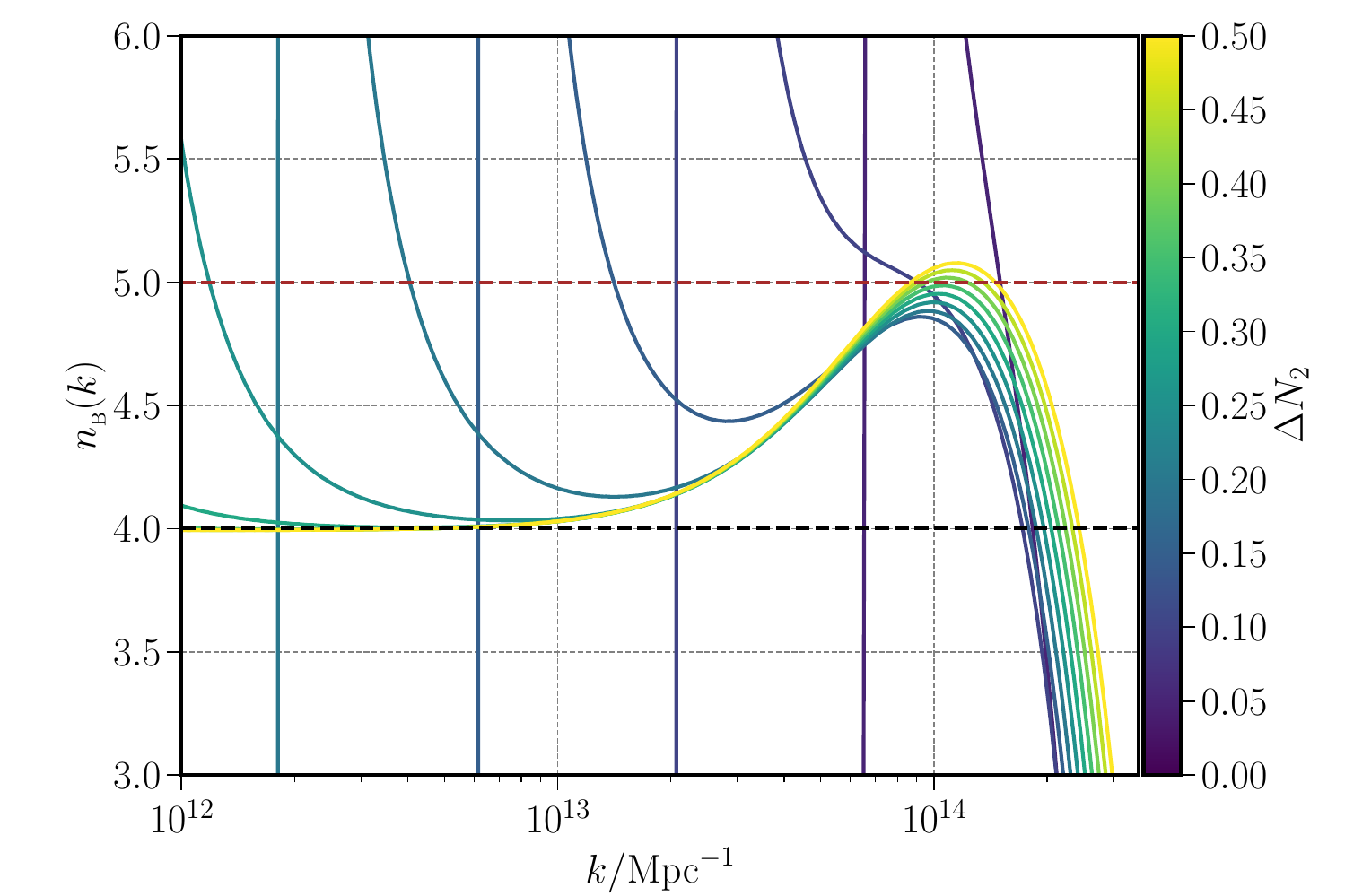}
\caption{\small The magnetic power spectrum $\pb(k)$ (in terms of $\pb^0$) and 
the associated spectral index $n_{\rm B}$ arising from the three-phase model of 
$J(N)$ are presented across variations of the parameters characterizing the intermediate
phase, namely $n_2$ and $\Delta N_2$.
We compute them in the parity-preserving case of $\gamma=0$ with a representative
value of $N_1=15$ (counted from the end of inflation).
For reference, we mark the value of $\pb(k) = \pb^0$ (in dashed black) in the plots of $\pb(k)$.
We examine the spectrum and $n_{\rm B}$ over the range of scales where the spectrum 
rises towards the peak, and we inspect the behavior across across a range of large negative 
values of $n_2 = [-20,-26]$ (on top panels).
For this analysis
we set the intermediate phase of $J$ to be brief yet finite with $\Delta N_2=0.5$. 
We find that the spectral index $n_{\rm B} \geq 5$ close to the peak.
We also examine the spectrum and $n_{\rm B}$ across a range of $\Delta N_2=[0,0.5]$, 
setting $n_2=-25$ (on bottom panels).
Once again, $n_{\rm B}$ easily reaches values $\geq 5$ for values of $\Delta N_2 \geq 0.4$.}
\label{fig:nB_numerical}
\end{figure}
Figure~\ref{fig:nB_numerical} shows that $n_{\rm B}$ can reach values as high as $5$ for $n_2 \in [-20,-26]$ and $\Delta N_2 \gtrsim 0.4$. This slightly exceeds the analytic maximum $n_{\rm B}=4.75$ obtained in~\cite{Atkins:2025pvg}. The discrepancy between the analytic and numerical maxima can be attributed to the finite duration $\Delta N_2$ of the intermediate phase. Notably, a steeper spectral growth is phenomenologically advantageous when confronting the model with observations, as it more readily evades CMB constraints~\cite{Durrer:2013pga}.

\subsection{Magnetic power spectrum - the parity-violating case}
\label{sec_sc1pv}

We proceed to study the three-phase model of $J$ with inclusion of the
parity-violation in the action. This is an important new ingredient of our work. See also \cite{Ng:2014lyb,Sharma:2018kgs,Okano:2020uyr,Brandenburg:2021bfx}.
In the first scenario  of this section we keep the
parameter $\gamma$
constant during inflation. Our fiducial values for the parameters are as follows.
The parity-violating term in the action is set to $\gamma=1$; 
the onset of intermediate phase to $N_1 = 15$ (counted from the end of inflation); 
the duration of the phase to  $\Delta N_2=3$; and the index to $n_2 = -3$. 
Note that the value of $n_2=-3$ is the dual to the case of $n_2=2$ as they both 
lead to the same solution for $\cAk$ when $\gamma=0$.
The choices of parameters are made to produce the peak in $\pb(k)$ over small scales 
of around $k \simeq 10^{14}\,\mpcinv$ and hence induce gravitational waves over 
corresponding frequencies of $f \simeq 0.1\,$Hz.
If this setup is realized in an ultra slow roll model suitable model of $J(\phi)$,
$N_1$ denotes the onset of ultra slow roll regime and our choice is consistent with
the corresponding constraint on it from observational data over large 
scales~\cite{Ragavendra:2024yfp}.
To understand the effect of these model parameters on $\pb^\lambda(k)$, we vary them 
around the aforementioned fiducial values and examine the associated implications.

\begin{figure}[t!]
\centering
\includegraphics[width=0.49\linewidth]{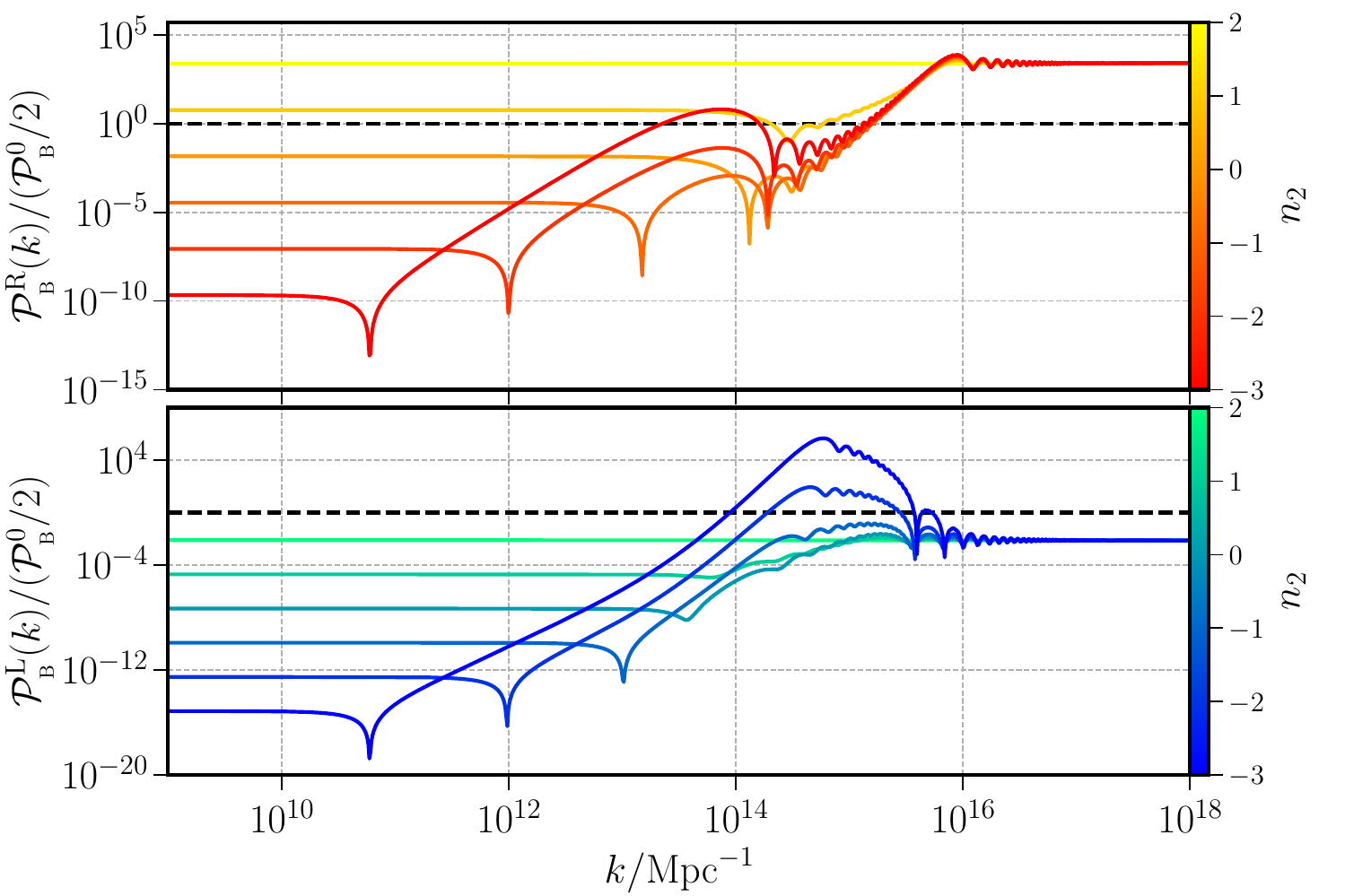}
\includegraphics[width=0.49\linewidth]{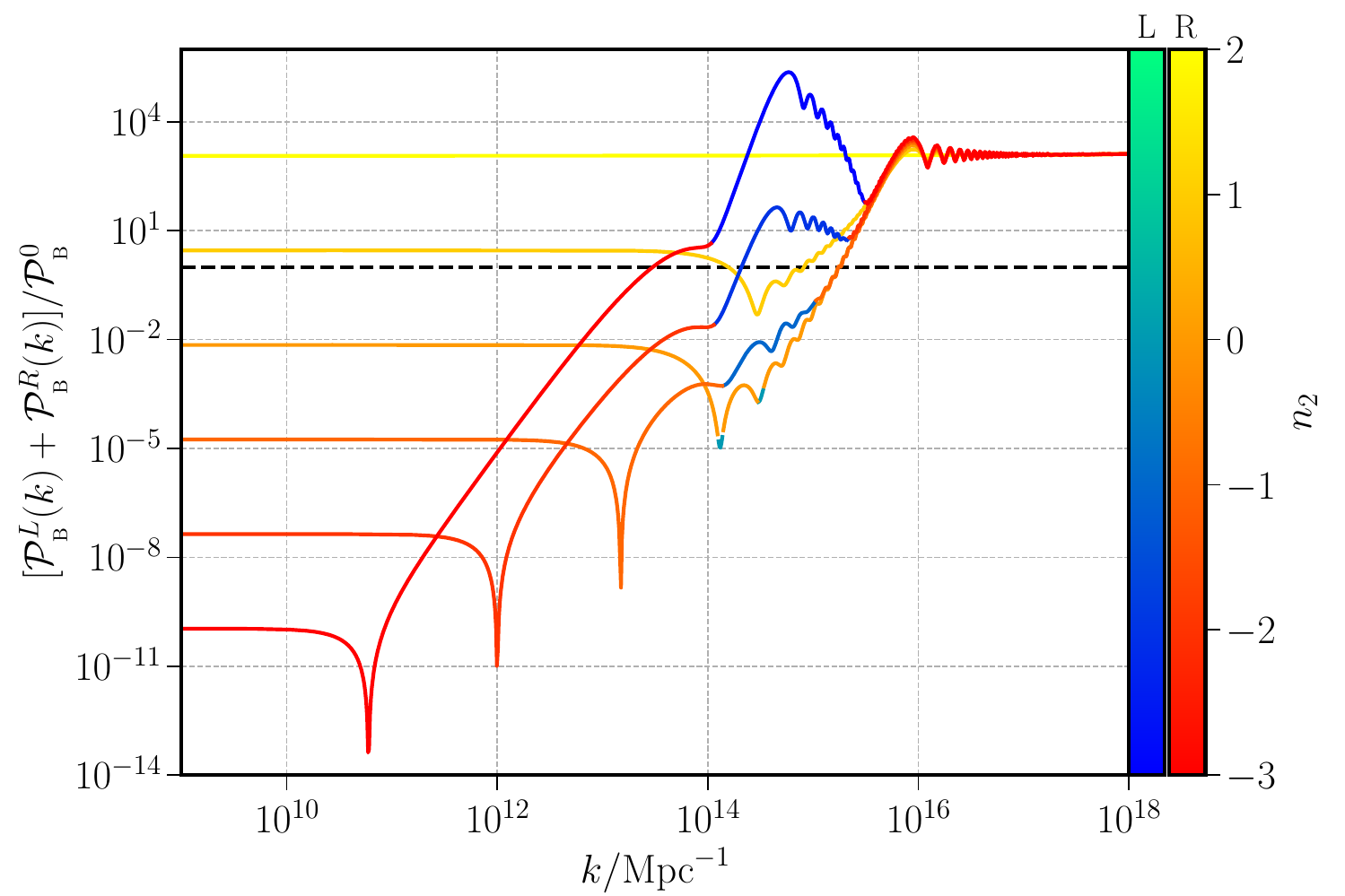}
\caption{\small We present $\pb^{\rm L}(k)/(\pb^0/2)$ and $\pb^{\rm R}(k)/(\pb^0/2)$ 
(on the left) and the total power spectrum $\sum_\lambda\pb^\lambda(k)/\pb^0$ 
(on the right) across a range of $n_2=[-3,2]$.
For reference, we also plot the spectrum of the non-helical case without the 
intermediate phase of $J$ where $\pb^{\rm L}=\pb^{\rm R}=\pb^0/2$ (in dashed black).
We have chosen $\gamma=1$, $N_1=15$ and $\Delta N_2=3$ in the functional form 
of $J(N)$.
This choice of parameters lead to the wavenumber corresponding to the onset of
the intermediate phase of $J$ to be $k_1=9.92\times 10^{13}\,\mpcinv$.
For $n_2=2$, we obtain $\pb^{\rm L}(k)$ and $\pb^{\rm R}(k)$ to be 
scale-invariant but with a significant difference in magnitude due to non-zero 
$\gamma$.
With the decrease in $n_2$, the features of dip, rise and peak become 
prominent in both the spectra.
The total power spectrum (on the right) is predominantly right-circular but for 
a brief window of scales between the asymptotic values, where the power increases and turns left-circular in polarization with decrease in $n_2$.
Hence, for $n_2=-3$ with $\Delta N_2=3$, we obtain an interesting double-peaked 
spectrum with distinct polarization for each peak.}
\label{fig:pb-n2-helical}
\end{figure}

We begin by inspecting the behavior of $\pb^\lambda(k)$ for each helicity against 
the variation of $n_2$.
In Fig.~\ref{fig:pb-n2-helical}, 
we present the spectra $\pb^{\rm L, R}(k)$ computed numerically, and the total power spectrum
$\pb(k) = \pb^{\rm L}(k) + \pb^{\rm R}(k)$ 
across a range of choices for $n_2$. 
The spectrum  is  scale-invariant and right-circular for $n_2=2$.
As we vary $n_2 = [-3,2]$, the features of dip, rise and peak become prominent 
in $\pb^\lambda(k)$ of both helicities. 

\smallskip

However, due
to the parity-violating term in the action, the behavior around the peak are different for the two helicities. 
Although the overall power in right-circular mode is enhanced, and 
the left-circular mode  suppressed --  as expected with $\gamma\neq 0$ --  the 
trend  is {\it reversed} around the scales approaching the peak of the spectrum.
This phenomenon leads to a double-peaked structure of total magnetic power, characterized by  altering 
helicities. We interpret it as
 due to interplay of democratic enhancement in power in both helicities due
to the intermediate phase of $J(N)$ and a preferential enhancement arising from 
the parity-violating term. 
Since for values of $n_2\neq 2$ the right-circular mode is suppressed with respect to the left-circular
mode around the peak,
the prominent peak of the power spectrum is left-circular in an otherwise 
right-circular spectrum.
We shall inspect this feature more closely in the next Section~\ref{sec_chfl}.  See also \cite{Talebian:2025jeg} for
related results in a similar context.

\begin{figure}[t!]
\centering
\includegraphics[width=0.49\linewidth]{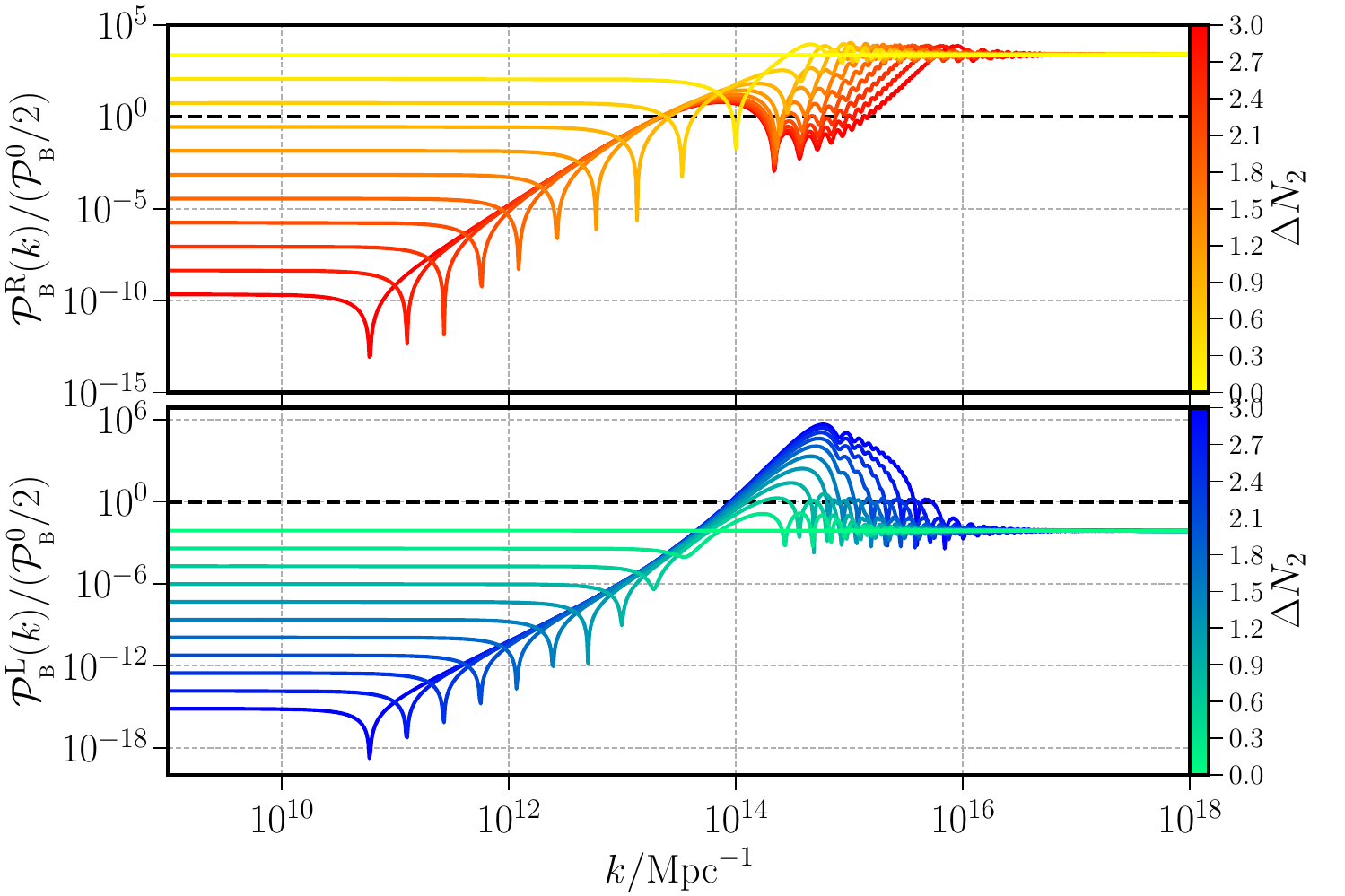}
\includegraphics[width=0.49\linewidth]{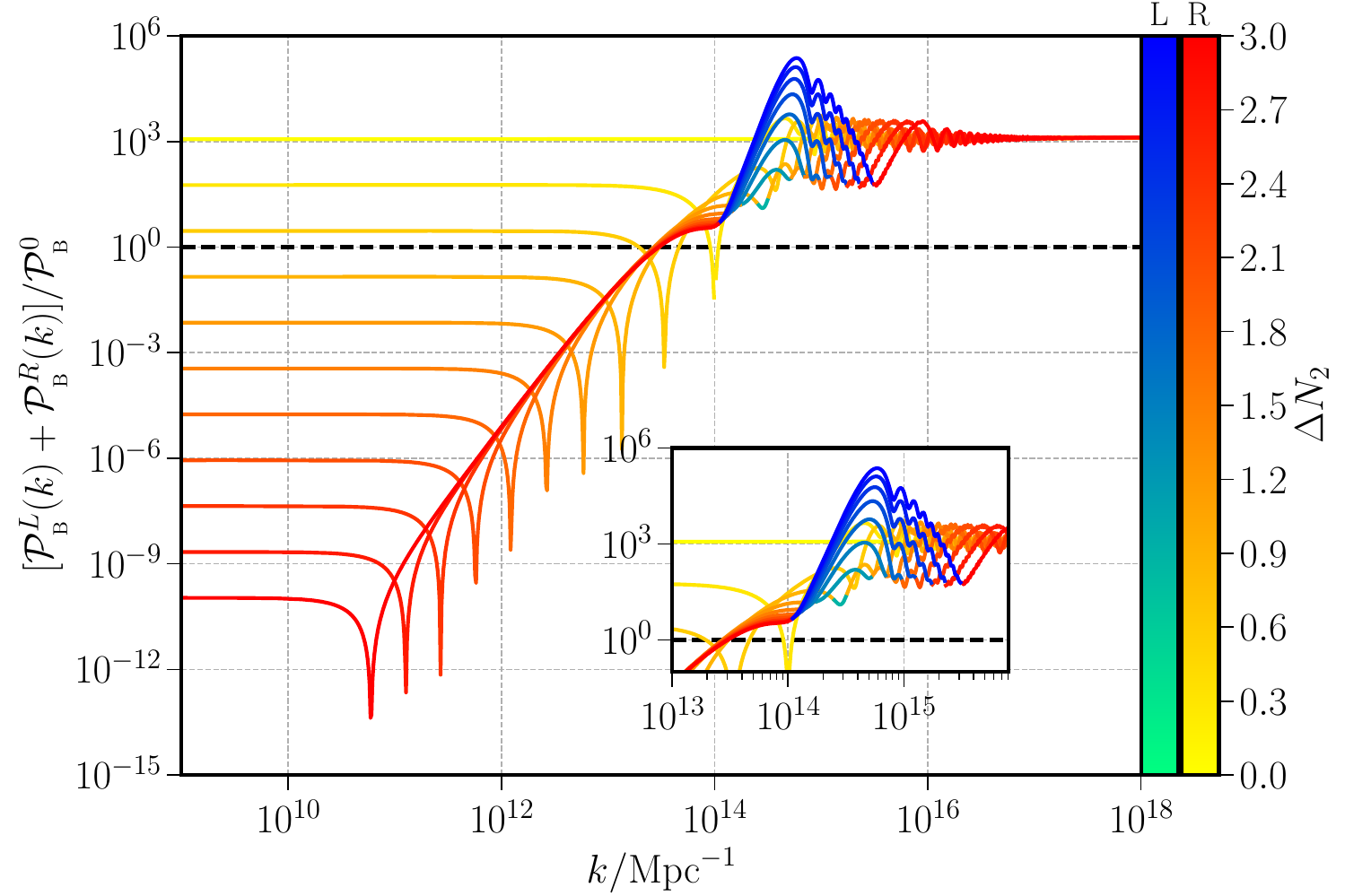}
\caption{\small We present $\pb^{\rm L}(k)$ and $\pb^{\rm R}(k)$ (on the left) 
and the total power spectrum $\sum_\lambda\pb^\lambda(k)$ (on the right) across 
a range of $\Delta N_2=[0,3]$, in terms of $\pb^0$ as in Fig.~\ref{fig:pb-n2-helical}.
We also plot the reference spectrum of $\pb^{\rm L}=\pb^{\rm R}=\pb^0/2$ 
(in dashed black).
We have chosen $\gamma=1$, $N_1=15$ and $n_2=-3$ in the functional form of $J(N)$.
For $\Delta N_2=0$, we obtain $\pb^{\rm L}(k)$ and $\pb^{\rm R}(k)$ to be 
scale-invariant but with $\pb^{\rm R}(k)$ dominating over $\pb^{\rm L}(k)$.
With the increase in $\Delta N_2$, the features of dip, rise and peak become 
prominent in both the spectra.
Around the peak of the spectra, $\pb^{\rm L}(k)$ begins to dominate over 
$\pb^{\rm R}(k)$ with increase in $\Delta N_2$. 
The total power spectrum (on the right) is predominantly right-circular but for
the window of scales that is peaked and left-circular in polarization (as focussed
in the inset).}
\label{fig:pb-deln2-helical}
\end{figure}

We then inspect the effect of the duration of the intermediate phase $\Delta N_2$ 
on the features $\pb(k)$.
We set $n_2=-3$ and $\gamma=1$, and present the results in Fig.~\ref{fig:pb-deln2-helical}.
Once again, as we vary $\Delta N_2=[0,3]$, we find the features of dip, rise
and peak becoming prominent in $\pb^\lambda(k)$ for both helicities.
Moreover, the altering helicity at the peak of the spectrum also becomes 
prominent with increase in $\Delta N_2$.

\begin{figure}[t]
\centering
\includegraphics[width=0.48\linewidth]{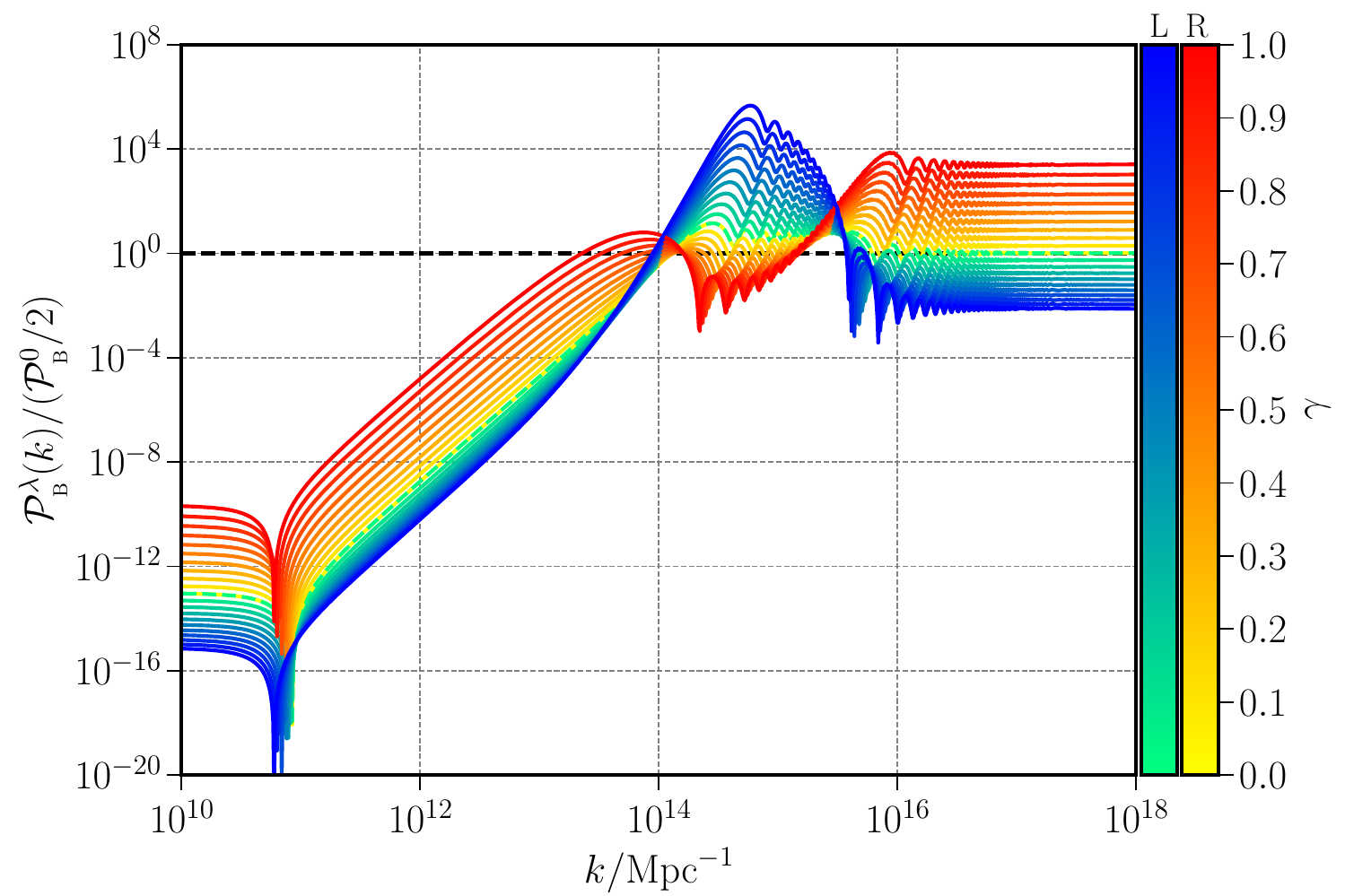}
\includegraphics[width=0.48\linewidth]{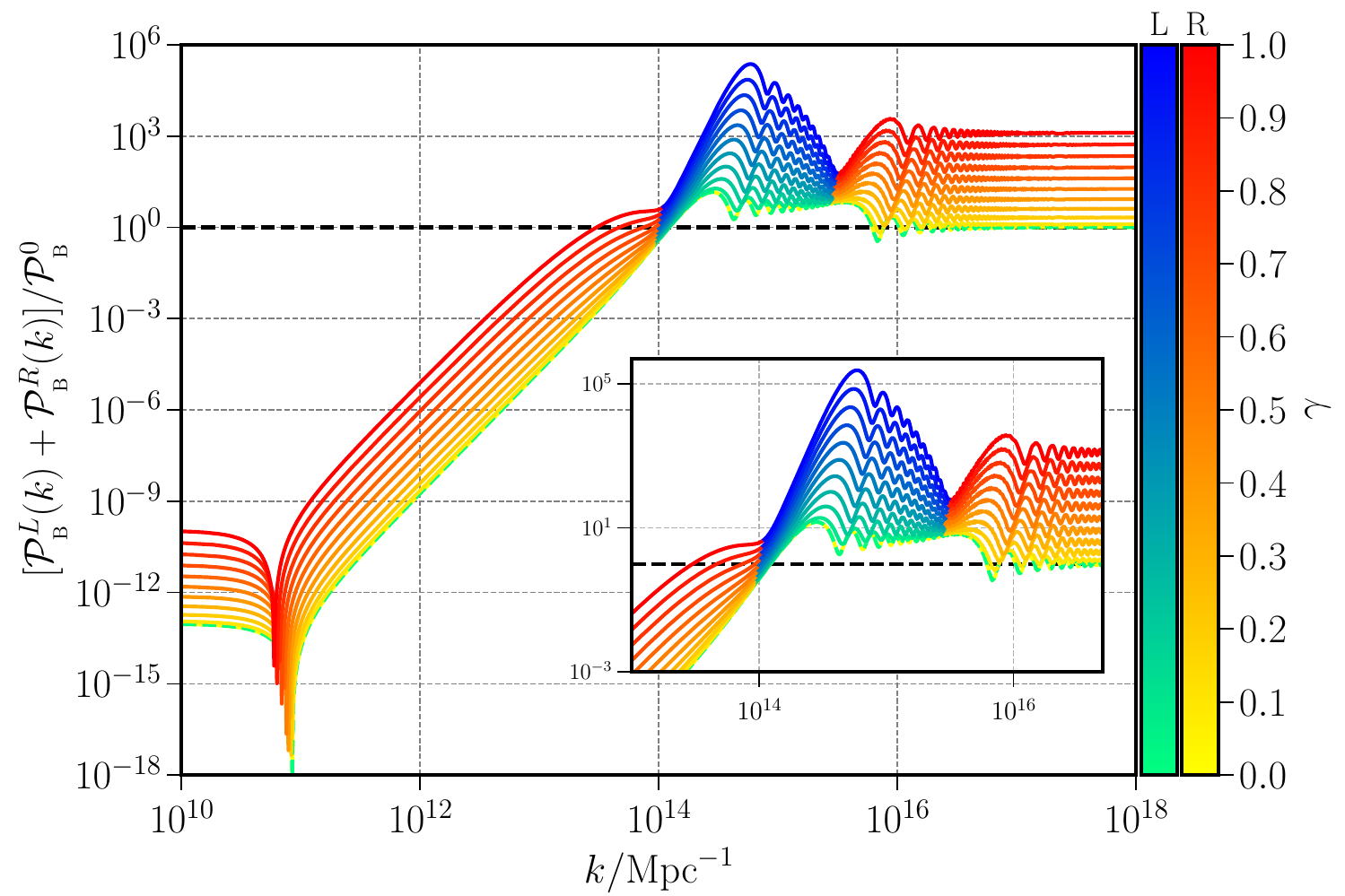}
\caption{\small We present the magnetic power spectrum $\pb^\lambda(k)$ for a range of 
$\gamma=[0,1]$ for individual helicities (on the left) and for the sum of both 
helicities (on the right) in terms of $\pb^0$ as in Figs.~\ref{fig:pb-n2-helical}
and~\ref{fig:pb-deln2-helical}.
The parameters of the model are set to be $N_1=15, n_2=-3,\, \Delta N_2=3$.
Evidently, for $\gamma=0$, we have both left and right helical modes having the 
same spectra.
The inset on the right panel zooms in on the range of scales around the peak 
that shall be relevant for the behavior of the spectra of induced GW.
Note that the case of $\gamma=1$ generates an amplification of about $10^5$ compared 
to the non-helical case of $\gamma=0$ at the peak of the spectrum.}
\label{fig:pb-vs-gamma}
\end{figure}

Lastly, we repeat our analysis varying the constant parameter $\gamma$ within 
the interval $\gamma=[0,1]$.
We set $n_2=-3$ and $\Delta N_2=3$ and present the spectra in Fig.~\ref{fig:pb-vs-gamma}.
We observe that for $\gamma=0$, $\pb^{\rm L}(k) = \pb^{\rm R}(k)$ with features of dip,
rise and peak in the spectra, as described in~\cite{Atkins:2025pvg}.
For values of $\gamma > 0$, we notice an  overall enhancement of $\pb^R(k)$ and a
suppression of $\pb^{\rm L}(k)$, except around the peak.
Around the peak, in fact,  we have a reversal of this trend, as noted earlier.

Besides, in the non-helical case of our setup ($\gamma=0$),
 the spectrum of electric field at the end of inflation
$\pe^\lambda(k)$ behaves as $\pe^\lambda(k) \simeq \pb^\lambda(k)(k/k_{\rm e})^2$,
with $k_{\rm e}$ corresponding to the wavenumber  at the end of inflation.
It turns out to be $\pe^\lambda(k) \simeq \pb^\lambda(k)$ when we turn on
$\gamma$ and becomes $\pe^\lambda(k) = \pb^\lambda(k)$ when $\gamma=1$, 
consistently mimicking the features in the spectrum, including alteration of helicity.
However, the electric field gets shorted during reheating with the production of charged particles and the associated high conductivity of the epoch~\cite{Subramanian:2009fu,Durrer:2013pga}.

Before proceeding further, we shall briefly remark about the observational bounds on 
the magnetic field strength over large scales in the current universe and the compatibility
of our model with the bounds.
The magnetic field strength over scales of $\lambda \geq 1\,{\rm Mpc}$ is bounded on the
lower end as $B_{\rm 1Mpc}>10^{-17}\,{\rm G}$,
from the $\gamma$-ray observations of extra-galactic sources~\cite{MAGIC:2022piy,Burmeister:2025lgo}.
On the other hand, it receives an upper bound from CMB anisotropies as
$B_{\rm 1Mpc}<10^{-9}\,{\rm G}$~\cite{Zucca:2016iur,Paoletti:2018uic}.
To examine these bounds in our model, we inspect the behavior of $\pb(k)$ over large scales
of CMB.

We can see that our model consistently leads to a scale-invariant $\pb(k)$ over CMB scales.
Further, the amplitude $\pb(k)\ll \pb^0$ over such scales, compared to $\pb^0$ over small scales.
For the fiducial values of $n_2=-3$ and $\Delta N_2=3$, with $\gamma=0$, we obtain
$\pb(k)\simeq 10^{-13}\,\pb^0$ over CMB scales.
We may relate the present day magnetic field strength smoothened over a given scale $\lambda$, 
denoted as $B_\lambda$, to $\pb(k)$ as~\cite{Zucca:2016iur,Paoletti:2018uic,Paoletti:2019pdi}
\begin{align}
B_\lambda^2 =& \left(\f{a_{\rm e}}{a_0}\right)^4
\int \f{\d\ln k}{4\pi}\,e^{-k^2\lambda^2}\pb(k)\,,
\end{align}
where $a_{\rm e}$ is the scale factor at the end of inflation and $a_0$ is the
scale factor today. 
Since we are interested in estimating $B_\lambda$ for $\lambda>1\,{\rm Mpc}$,
the field strength is assumed to follow the behavior of $1/a^2$ since the end of inflation, 
without any effects of turbulence.
Hence the ratio of scale factors in the above expression.
Using conservation of entropy, the ratio can be expressed in terms of the 
Hubble parameter during inflation as~\cite{Tripathy:2021sfb,Tripathy:2022iev}
\begin{align}
\f{a_0}{a_{\rm e}} =& 2.8\times 10^{28}
\left(\f{H}{10^{-5}\,\Mpl}\right)^{1/2}.
\end{align}
We have worked with Hubble parameter to be $H \simeq 7\times 10^{-6}\,\Mpl$, 
consistent with the bound on tensor-to-scalar ratio from CMB anisotropies.
Using these values, we may estimate the magnetic field strength arising from our model 
to be about $B_{\rm 1Mpc} \simeq 10^{-15}$\,G.
For $\gamma=1$, we get $\pb(k)\simeq 10^{-10}\,\pb^0$ over CMB scales
and hence $B_{\rm 1Mpc} \simeq 3\times 10^{-14}$\,G.
These values are within the existing bounds on the parameter
$10^{-17}{\rm G}< B_{\rm 1Mpc} <10^{-9}$G~\cite{Zucca:2016iur,Paoletti:2018uic,Paoletti:2019pdi,MAGIC:2022piy,Burmeister:2025lgo}.

\subsection{Altering helicities at the peak - an analytical understanding}
\label{sec_chfl}

The curious feature of altering helicities around the peak value of $\pb(k)$ warrants
closer examination and we may understand this effect in the following way.
Recall that the equation of motion governing $\cAk$ is Eq.~\eqref{eq:cAk-helical}
\begin{eqnarray}
\cAk^{\lambda\,\prime\prime} + \left(k^2 +2\xi^\lambda\gamma\,k\f{J'}{J} - \f{J''}{J}\right)\cAk^\lambda &=& 0\,.
\end{eqnarray}
During the phase when $J \propto a^2 \propto \tau^{-2}$ we get  $J'/J = \vert2/\tau\vert$
and $J''/J = 6/\tau^2$. 
Hence, the equation becomes
\begin{eqnarray}
\cAk^{\lambda\,\prime\prime} + \left(k^2 +4\xi^\lambda\gamma\,\left\vert\f{k}{\tau}\right\vert 
- \f{6}{\tau^2}\right)\cAk^\lambda &=& 0\,.
\end{eqnarray}
The parity-violating term enhances ${\cal A}^R_k$ (with $\xi^R=-1$) and suppresses 
${\cal A}^L_k$ (with $\xi^L=1$) for wavenumbers exiting the Hubble radius, i.e.
$\vert k\tau\vert \simeq \sqrt{6}$ during this regime.
The behavior of the enhanced mode in the asymptotic regime of $\vert k\tau\vert \ll 1$ is~\cite{Caprini:2014mja}
\begin{eqnarray}
\cAk^R &\propto & \f{e^{2\pi\gamma}}{(2\gamma)^{3/2}}\,.
\end{eqnarray}

On the other hand, in the phase when $J\propto a^{-3}\propto \tau^3$ 
we get $J'/J = -\vert3/\tau\vert$ while $J''/J = 6/\tau^2$.
This leads to the equation of motion being
\begin{eqnarray}
\cAk^{\lambda\,\prime\prime} + \left(k^2 -6 \lambda\gamma\left\vert\f{k}{\tau}\right\vert 
- \f{6}{\tau^2}\right)\cAk^\lambda &=& 0\,.
\end{eqnarray}
The difference in sign of $J'/J$ with respect to the earlier case leads to opposite 
effect on helicities.
The parity-violating term now enhances ${\cal A}^L_k$ and suppresses ${\cal A}^R_k$ for
modes attaining $\vert k\tau\vert \simeq \sqrt{6}$ during this regime.
The behavior of the enhanced mode in the asymptotic regime of $\vert k\tau\vert \ll 1$ is
\begin{eqnarray}
\cAk^L &\propto & \f{e^{3\pi\gamma}}{(3\gamma)^{3/2}}\,.
\end{eqnarray}
Thus we have different helicities enhanced over wavenumbers exiting Hubble radius 
in different regimes of $J(\tau)$.

Note that the enhancement of the mode function exiting Hubble radius during the phase of 
$J \propto a^{-3}$ is larger by a factor of $3/2$ in the exponent compared to the mode 
function exiting in the phase of $J \propto a^2$, i.e.
\begin{align}
{\cAk^L}_{(J\sim a^{-3})} &\propto \left[{\cAk^R}_{(J\sim a^2)}\right]^{3/2}\,.
\end{align}
Hence the corresponding power spectra for the two cases may be related as
\begin{align}
\pb^L(k)_{(J\sim a^{-3})} &\propto \left[\pb^R(k)_{(J\sim a^2)}\right]^{3/2}\,.
\end{align}
This relation explains the difference between enhancements of amplitudes of power over
large scales and the power at the peak.
For $\gamma = 1$, over large scales $\pb^L(k \ll k_1)$ is enhanced by about $10^3$ compared
to the case of $\gamma=0$.
The value $\pb^R(k \simeq k_1)$   around the peak is enhanced by about 
$10^{9/2} \simeq 3\times 10^5$.


\subsection{Backreaction}

The issue of backreaction may become crucial in the presence of parity-violating 
interactions, which lead to chiral instabilities~\cite{Durrer:2010mq,Urban:2011bu,Bartolo:2015dga,Sharma:2018kgs,Chowdhury:2018mhj,Tripathy:2021sfb}.
To trust our set-up, we must ensure that the energy density of the electromagnetic field 
$\rho_{_{\rm EM}}$ remains sub-dominant to the background energy density driving inflation 
$\rho_{_{\rm I}}$.
In case of magnetogenesis with monotonic evolution of the coupling function $J$,
value of $\gamma \gtrsim 2.5$ is shown to cause strong backreaction~\cite{Tripathy:2021sfb}. Interestingly, we {\it do not encounter} issues with  backreaction in our scenario, making it a viable model for chiral magnetogenesis from inflation without issues related
with backreaction. 

In fact, let us examine   the case of a three-phase evolution of $J(N)$ in the scenario we are considering here.
Recall that the tensor power spectrum $\pt(k) \simeq 2H^2/\Mpl^2$. 
Therefore the background energy density during inflation is $\rho_{_{\rm I}}(\tau)$
\begin{equation}
\rho_{_{\rm I}}(\tau) = 3H^2\Mpl^2 \simeq \Mpl^4\,\pt(k).
\end{equation}
On the other hand,  the electromagnetic energy density is 
\begin{equation}
\rho_{_{\rm EM}}(\tau) = \f{1}{2}\int \d \ln k\,[\pb(k) + \pe(k)] \simeq \int \d\ln k\,\pb(k)\,,
\end{equation}
where the last expression is justified in case of $\gamma={\cal O}(1)$ leading to
$\pb(k) \simeq \pe(k)$.
Hence, $\pb(k)/(\pt(k)\Mpl^4)$ is an equivalent estimate of $\rho_{_{\rm EM}}/\rho_{_{\rm I}}$.
Ensuring this ratio to remain below unity across scales indicates control over
backreaction.

\begin{figure}[t!]
\centering
\includegraphics[width=0.5\linewidth]{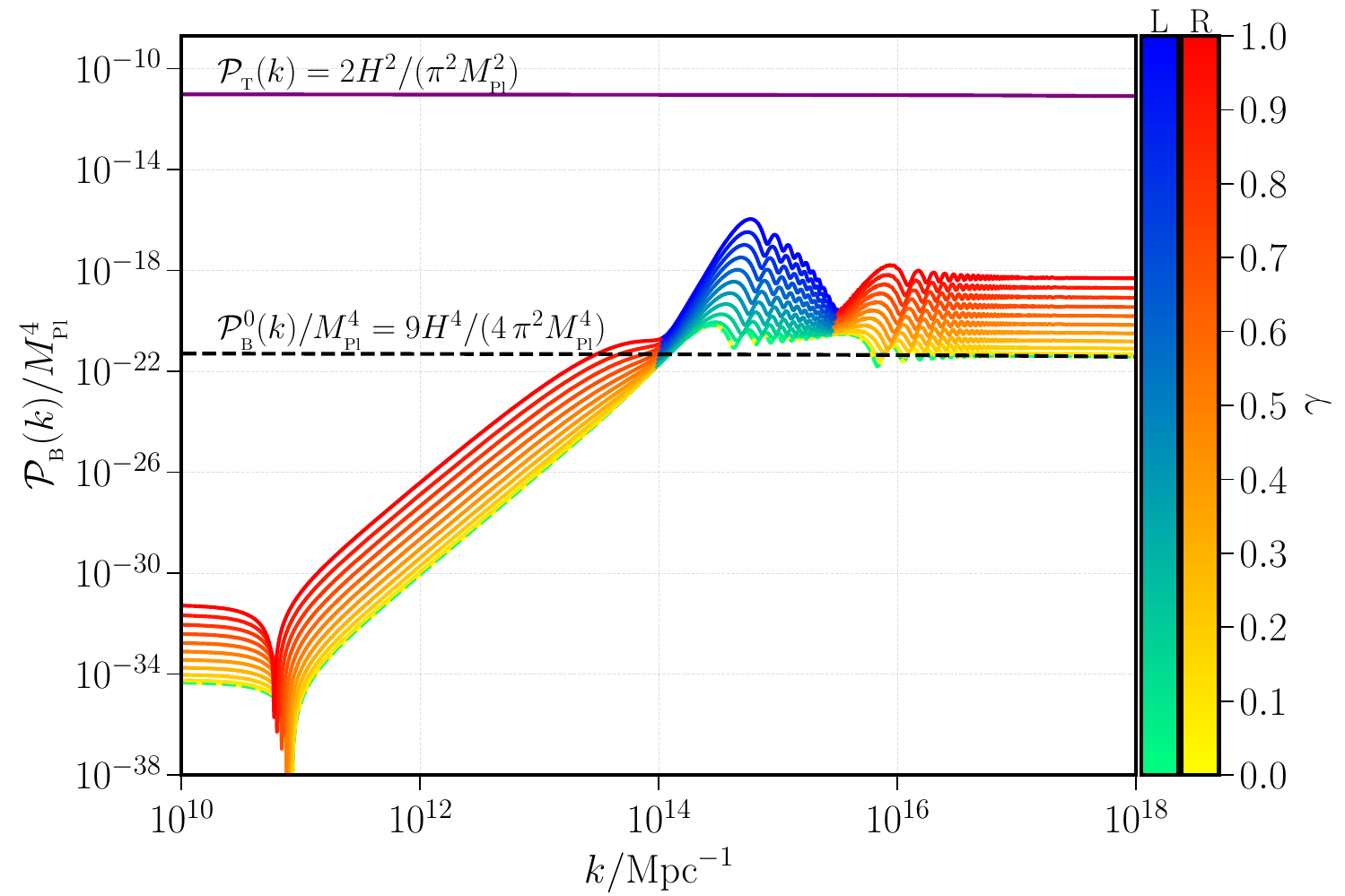}
\caption{\small The total power spectrum $\pb(k)=\sum_\lambda\pb^\lambda$ is presented 
in units of $\Mpl^4$ against the power spectrum of primary tensor perturbations 
$\pt(k)$, for a range of $\gamma=[0,1]$.
We set the model parameters to be $n_2=-3,\,N_1=15$ and $\Delta N_2=3$.
The plot illustrates that the ratio of $\pb(k)/(\pt(k)\Mpl^4)$ is small even at 
the peak amplitude of $\pb(k)$.
At the asymptotic value over small scales, this ratio is approximately
$\pb(k)/(\pt(k)\Mpl^4) \simeq f(\gamma)(H/\Mpl)^2$, where $f(\gamma)$ is typically
${\cal O}(10^3)$ for $\gamma=1$.
So, as long as $H/\Mpl \leq 10^{-3/2}$, the backreaction due to electromagnetic 
energy density is small in the model of interest.}
\label{fig:pb-vs-pt}
\end{figure}

We illustrate the comparison of $\pb(k)$ against $\Mpl^4\pt(k)$ in Fig.~\ref{fig:pb-vs-pt}
over the range of values of $\gamma$ we work with. (We choose $H \simeq 7\times 10^{-6}\Mpl$, so to be consistent with bounds on the tensor to scalar
ratio over CMB scales.) 
The spectrum $\pb(k)$ is suppressed from $\pb^0=9H^4/4\pi^2$ over large scales, 
and so it is very small compared to $\Mpl^4\pt(k)$.
Crucially, the ratio of $\pb(k)/(\pt(k)\Mpl^4)$ is small at the peak amplitude 
of $\pb(k)$, even for the maximum value of $\gamma=1$.
Over small scales, where $\pb(k)$ settles to the asymptotic value of $\pb^0$, 
this ratio becomes $\pb(k)/(\pt(k)\Mpl^4) \simeq f(\gamma)(H/\Mpl)^2$, 
where $f(\gamma)$ is ${\cal O}(10^3)$ for 
$\gamma=1$~\cite{Sharma:2018kgs,Chowdhury:2018mhj,Tripathy:2021sfb}.
So --  as long as $H/\Mpl \leq 10^{-3/2}$ -- thanks to the fact that the magnetic field is enhanced rapidly at small scales only, we conclude that the backreaction due to electromagnetic 
field is negligible, as stated above.

\section{Second scenario: analytical results}
\label{sec_scen2}
\label{sec_analytical}

In this section we present a second scenario for non-slow-roll magnetogenesis. We extend the analytical methods of \cite{Atkins:2025pvg} to a setup including parity violation, with the aim of obtaining analytic insight into the numerical results presented in Section~\ref{sec_scen1}. The approach closely follows the techniques of \cite{Tasinato:2020vdk,Tasinato:2023ukp}, adapted to our context; differences and clarifications are discussed where relevant.

The starting point is the evolution equation for the helical modes during de Sitter inflation -- Eq.~\eqref{eq:cAk-helical} which we report here:
\begin{eqnarray}
\cAk^{\lambda\,\prime\prime}(\tau) + \left(k^2 +2 \lambda k \,\gamma(\tau)\frac{J'(\tau)}{J(\tau)} - \frac{J''(\tau)}{J(\tau)}\right)\cAk^\lambda(\tau) &=& 0 \,,
\label{eq:cAk-helicalV2}
\end{eqnarray}
with $\lambda=\pm1$, and where we allow $\gamma(\tau)$ to be time dependent.

As a reference, in the standard case in which $J(\tau)\propto a^2(\tau)$ and $\gamma(\tau)=0$ during inflation, with $a(\tau)=-1/(H_I \tau)$, the solution connected
to the Bunch-Davies vacuum is the well-known mode
function
\begin{equation}
\label{eq_stmf}
{\cal A}_k(\tau)=\frac{3\,a^2(\tau)\,H_I^2}{\sqrt{2}\,k^{5/2}}\,e^{-i k \tau}
\left(1+ i k \tau-\frac{k^2 \tau^2}{3} \right).
\end{equation}
We are interested in a  general class of models in which
\begin{equation}
J(\tau)=a^2(\tau)\,\sqrt{\omega(\tau)},
\end{equation}
with $\omega(\tau_R)=1$ at the end of inflation. We assume the following time dependence for the functions $J$ and $\gamma$:
\begin{eqnarray}
\omega(\tau)&=&
\begin{cases}
\omega_0 \quad & \text{for } \tau<\tau_1, \\[3pt]
\text{continuous but rapidly varying} \quad & \text{for } \tau_1<\tau<\tau_2,\\[3pt]
1 \quad & \text{for } \tau_2<\tau<\tau_R,
\end{cases}
\\[6pt]
\gamma(\tau)&=&
\begin{cases}
0 \quad & \text{for } \tau<\tau_1, \\[3pt]
\text{continuous but rapidly varying} \quad & \text{for } \tau_1<\tau<\tau_2,\\[3pt]
0 \quad & \text{for } \tau_2<\tau<\tau_R.
\end{cases}
\label{ans_gam}
\end{eqnarray}
The two times $\tau_1$ and $\tau_2$ are close to each other
-- in fact we assume
\begin{equation}
\label{eq_shorint}
\frac{\tau_2-\tau_1}{\tau_1}\ll 1,
\end{equation}
so that the interval $\tau_1<\tau<\tau_2$ is short: within this brief period $J(\tau)$ may significantly deviate from a power law and parity-violating effects encoded in $\gamma(\tau)$ can be active. (Later in this section, we consider the limit
of Eq.~\eqref{eq_shorint} infinitesimally small.)

Physically, as described in the previous sections, this situation can be realised in inflationary models where the inflaton kinetic energy (or $\dot\phi$) changes rapidly for a short time.
In such scenarios one can envisage set-up with  $\gamma\sim\dot\phi^2$, so that $\gamma$ is negligible during slow roll and becomes nonzero (and potentially large) only during the brief non-slow-roll phase. Notice that the choice of profile
in Eq.~\eqref{ans_gam} is different with respect to what was discussed in Section~\ref{sec_sc1pv},
where we instead assume $\gamma=1$ throughout inflation.

To obtain full analytic solutions 
for this system we adopt the following Ansatz for the mode function,
\begin{eqnarray}
\label{eq_ans2}
{\cal A}_k(\tau)&=&\frac{3\,a^2(\tau)\,\sqrt{\omega(\tau)}\,H_I^2}{\sqrt{2}\,k^{5/2}}
e^{-i k \tau}\Big[1+ i k \tau - \frac{k^2 \tau^2}{3}
-k \tau_0 \Gamma_1(\tau)\nonumber\\
&&\qquad\qquad +(i k \tau_0)^2\big(G_{(2)}(\tau) +i \Gamma_{(2)}(\tau)\big)
+(i k\tau_0)^3\big(G_{(3)}(\tau) +i \Gamma_{(3)}(\tau)\big) + \dots \Big],
\end{eqnarray}
where $\tau_0$ is a pivot quantity whose explicit value will cancel out of final observables. The solution is constructed piecewise, extending
the methods of \cite{Byrnes:2018txb}:
\begin{enumerate}
\item For $\tau<\tau_1$ we recover the standard solution ($J=a^2$, $\gamma=0$) so all $G_i,\Gamma_j$ vanish
and we reduce to Eq.~\eqref{eq_stmf}.
\item For $\tau_1<\tau<\tau_2$ we determine the functions $G_i,\Gamma_j$ by inserting the Ansatz \eqref{eq_ans2} into Eq.~\eqref{eq:cAk-helicalV2} and solving order-by-order in powers of $k$.
\item For $\tau_2<\tau<\tau_R$ we again have the standard slow-roll form ($J=a^2,\gamma=0$) and we impose Israel matching conditions at $\tau=\tau_2$ to fix the post-intermediate-stage coefficients.
\end{enumerate}
Working within the short interval $\tau_1<\tau<\tau_2$, we expand all quantities about $\tau_1$ and define $\Delta\tau\equiv\tau-\tau_1$ with $0\le\Delta\tau\ll\tau_1$. From the $k^1$ order we obtain the differential relation
\begin{equation}
\frac{d}{d\tau}\!\left[ \frac{\tau_0 \, \omega(\tau) \, \Gamma_1'(\tau)}{\tau^4}  \right]
= \lambda \,\gamma(\tau)\frac{d}{d\tau} \!\left[ \frac{ \omega(\tau) }{\tau^4} \right],
\end{equation}
which is formally integrated to
\begin{equation}
\label{fsol_ga1}
\tau_0 \Gamma_1(\tau)=\lambda\int_{-\infty}^\tau d \tau_a\,\frac{\tau_a^4}{\omega(\tau_a)}
\int_{-\infty}^{\tau_a} d \tau_b\,\gamma(\tau_b)\frac{d}{d\tau_b} \left[ \frac{ \omega(\tau_b) }{\tau_b^4} \right].
\end{equation}
We
introduce  the following parameters evaluated at $\tau=\tau_1$,
\begin{eqnarray}
\alpha&\equiv& \left( \frac{d \ln \omega}{d \ln \tau}\right)_{\tau=\tau_1},
\\
\beta&\equiv& \left( \frac{d  \gamma}{d \ln \tau}\right)_{\tau=\tau_1}.
\end{eqnarray}
Expanding the formal solution \eqref{fsol_ga1} in $\Delta\tau$ we find that the leading nonzero contribution to $\Gamma_1$ is cubic,
\begin{equation}
\tau_0 \Gamma_1(\tau)=\lambda\,\frac{\beta\,(\alpha-4)\,{\Delta \tau}^3}{6\tau_1^2} + \mathcal{O}(\Delta\tau^4).
\label{eq_G1_leading}
\end{equation}

From the term $k^2$ in the expansion  we  obtain two conditions, which once Taylor-expanded  in $\Delta \tau$ give 
\begin{eqnarray}
\tau_0^2 G_2(\tau)&=&\frac{\alpha}{6}\,\Delta \tau^2 + \mathcal{O}(\Delta\tau^3),
\\
\tau_0^2 \Gamma_2(\tau)&=&\lambda\frac{\beta (\alpha-4)}{6 \tau_1}\,\Delta \tau^3 + \mathcal{O}(\Delta\tau^4).
\end{eqnarray}
At $k^3$ order the leading results are
\begin{eqnarray}
\tau_0^3 G_3(\tau)&=& \frac{\alpha}{6}\,\tau_1\,\Delta \tau^2 + \mathcal{O}(\Delta\tau^3),
\\
\tau_0^3 \Gamma_3(\tau)&=&\lambda\frac{\beta (\alpha-4)}{18}\,\Delta \tau^3 + \mathcal{O}(\Delta\tau^4).
\end{eqnarray}
For higher orders $k^n$ with $n\ge4$ the coefficients obey simple recursion relations (for derivatives evaluated at $\tau_1$),
\begin{eqnarray}
\tau_0^n G_n^{(n-1)}(\tau_1)&=& 2\,\tau_0^{\,n-1}\,G_{n-1}^{(n-2)}(\tau_1),
\\
\tau_0^n \Gamma_n^{(n)}(\tau_1)&=& 2\,\tau_0^{\,n}\,\Gamma_{n-1}^{(n-1)}(\tau_1),
\end{eqnarray}
which imply
\begin{eqnarray}
\tau_0^n G_n^{(n-1)}(\tau_1)&=& 2^{\,n-3}\,\frac{\alpha \tau_1}{3},
\\
\tau_0^n \Gamma_n^{(n)}(\tau_1)&=& 2^{\,n-3}\,\lambda\,\beta \frac{(\alpha-4)}{3}.
\end{eqnarray}
Hence, keeping leading Taylor terms for each $n\ge3$,
\begin{eqnarray}
\label{res_gn}
\tau_0^n G_n(\tau)&=&2^{\,n-3}\,\frac{\alpha \tau_1}{3}\,\frac{\Delta \tau^{\,n-1}}{(n-1)!} + \dots,
\\
\label{res_gan}
\tau_0^n \Gamma_n(\tau)&=&2^{\,n-3}\,\lambda\,\beta \frac{(\alpha-4)}{3}\,\frac{\Delta \tau^{\,n}}{n!} + \dots .
\end{eqnarray}

Plugging all these results in Ansatz~\eqref{eq_ans2}, they lead to an exact resummation 
to exponentials. 
In fact, we use
\begin{eqnarray}
\sum_{n=3}^{\infty} (ik\tau_0)^n G_{(n)}(\tau)
&=& \frac{i k \tau_1\,\alpha}{12}\left( e^{2 i k \Delta\tau}-1 -2 i k \Delta\tau \right),
\\
i\sum_{n=3}^{\infty} (ik\tau_0)^n \Gamma_{(n)}(\tau)
&=& i\,\lambda\,\beta \frac{(\alpha-4)}{24}\left( e^{2 i k \Delta\tau}-1 -2 i k \Delta\tau+2\,k^2 \Delta\tau^2 \right).
\end{eqnarray}
To conclude,  
 the resulting mode function in the short interval $\tau_1<\tau<\tau_2$ can be written as
\begin{eqnarray}
{\cal A}_k(\tau)&=&\frac{3\,a^2(\tau)\,\sqrt{\omega(\tau)}\,H_I^2}{\sqrt{2}\,k^{5/2}}
e^{-i k \tau}\Big[1+ i k \tau - \frac{k^2 \tau^2}{3}
- \lambda\,\frac{\beta\,(\alpha-4)\, k \,{\Delta \tau}^3}{6\tau_1^2}\nonumber\\
&&\qquad
- k^2 \Big(\frac{\alpha}{6}\,\Delta \tau^2 +i \lambda\frac{\beta (\alpha-4)}{6 \tau_1}\,\Delta \tau^3\Big)
+\frac{i k \tau_1\,\alpha}{12}\big( e^{2 i k \Delta\tau}-1 -2 i k \Delta\tau\big)\nonumber\\
&&\qquad
+i\,\lambda\,\beta \frac{(\alpha-4)}{24}\big( e^{2 i k \Delta\tau}-1 -2 i k \Delta\tau+2\,k^2 \Delta\tau^2 \big)
\Big].
\label{eq_ans3}
\end{eqnarray}
This expression displays the explicit chirality dependence through $\lambda$ and captures the leading modifications arising from the short non-slow-roll episode.

Matching this intermediate solution to the final slow-roll phase using Israel junction conditions yields, for $\tau>\tau_2$, the standard form
\begin{eqnarray}
{\cal A}_k(\tau)=\frac{3\,a^2(\tau)\,H_I^2}{\sqrt{2}\,k^{5/2}}
\Big[
C_1 \left(1+ i k \tau - \frac{k^2 \tau^2}{3}  \right) e^{-i k \tau}
+C_2 \left(1- i k \tau - \frac{k^2 \tau^2}{3}  \right) e^{i k \tau}
\Big],
\label{eq_ans5}
\end{eqnarray}
where the  coefficients $C_{1,2}$ are fixed by the matching conditions of the function and its derivatives at $\tau_2$.

\begin{figure}[t!]
    \centering
    \includegraphics[width=0.47\linewidth]{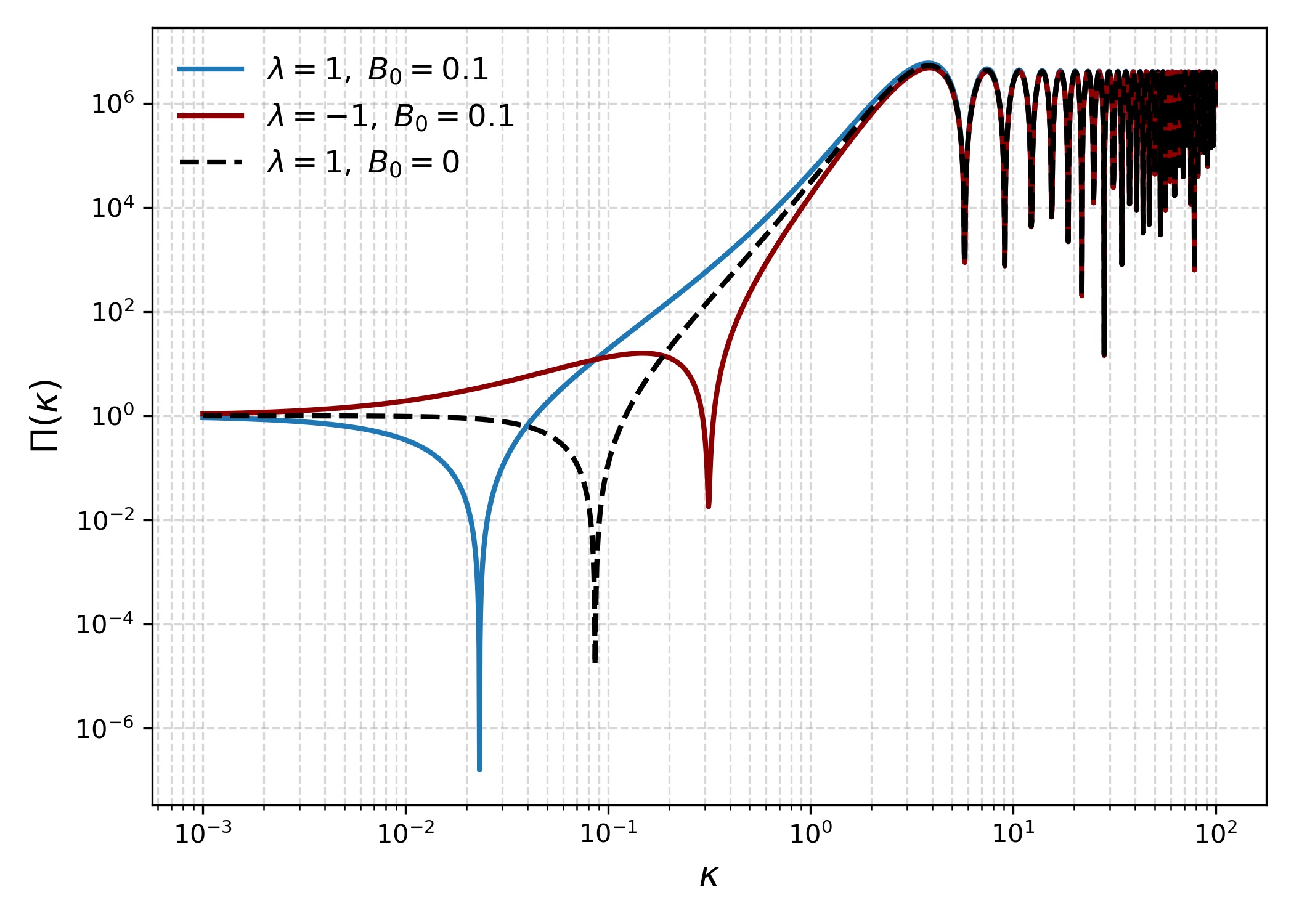}
    \includegraphics[width=0.47\linewidth]{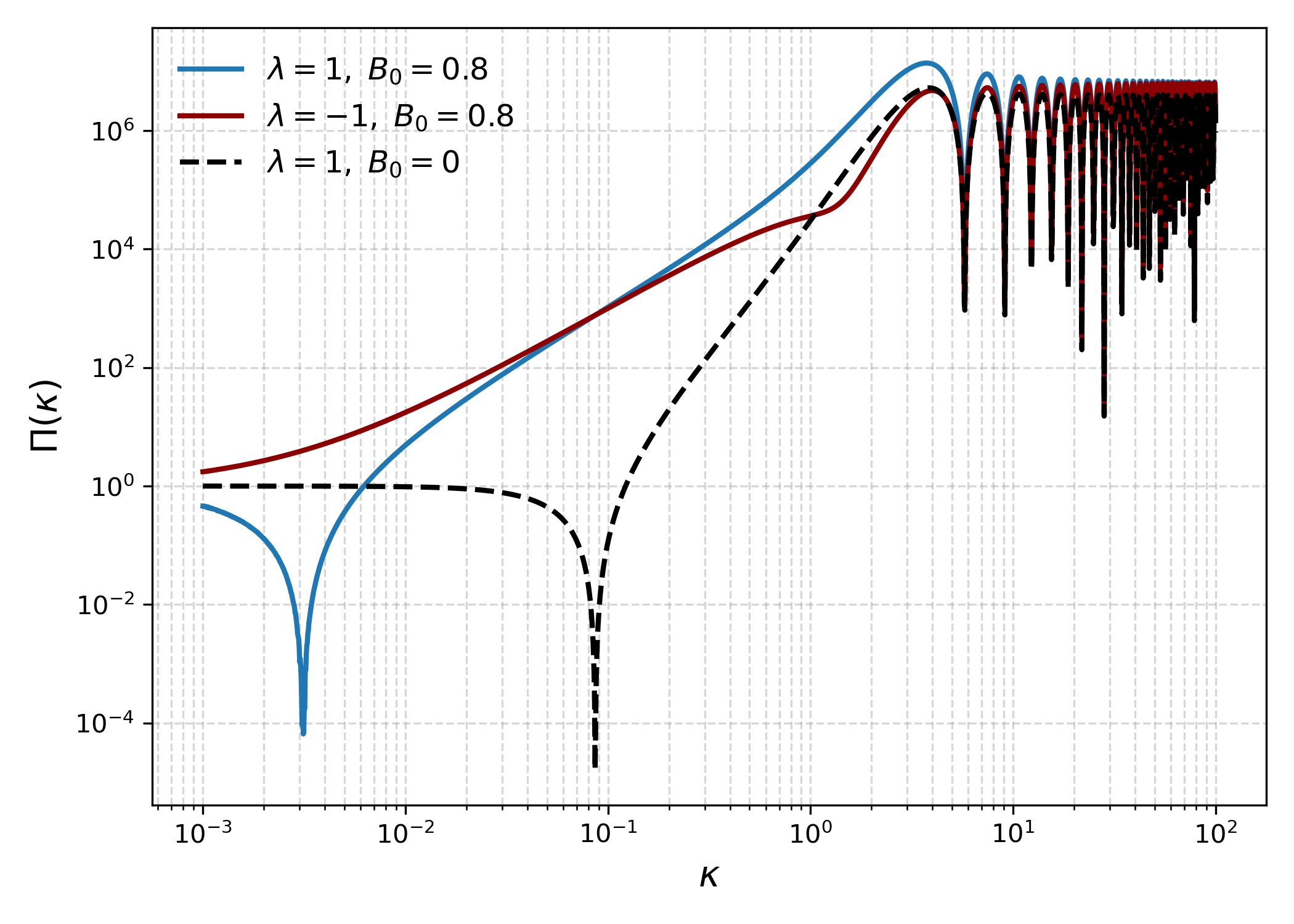}
    \caption{\small 
    Analytical spectra evaluated
    with the limiting
    formula \eqref{eq_sumC}.
    }
    \label{fig_simPB}
\end{figure}

We proceed making  additional steps that further simplify  our expressions for the 
mode functions.  We take the limits
\begin{eqnarray}
&&
\Delta \tau \to 0 \hskip0.5cm,\hskip0.5cm \alpha \to \infty
\hskip0.5cm,\hskip0.5cm \beta \to \infty
\nonumber
\\
&&
 \alpha  \Delta \tau  = 2 \Pi_0
 \hskip0.5cm,\hskip0.5cm    \Pi_0
{\beta}/{\alpha}\,=\,B_0
\end{eqnarray}
keeping
$\Pi_0$, $B_0$ finite. 
This is motivated by the  `t Hooft limit \cite{tHooft:1973alw} which
one encounters in particle physics. 
We work in terms
of the dimensionless quantity
\begin{equation}
\kappa= -k \tau_1
\end{equation}
with $\tau_1$ the conformal time during inflation when the non-slow-roll
phase occurs.
Then, 
the quantities $C_{1,2}$ appearing in Eq.~\eqref{eq_ans5}
reduce to
\begin{eqnarray}
C_1&=&
1 + \left(1 - \frac{3 i}{\kappa^3} - \frac{2 i}{\kappa}
- \frac{i B_0 \left(9 + 3\kappa^2 + \kappa^4\right)\lambda}{\kappa^4}\right)\Pi_0,
\\
C_2&=&
- \frac{e^{2 i \kappa} (i + \kappa)\,(-3 + 3 i \kappa + \kappa^2)\,\Pi_0}{\kappa^3}
+ \frac{i B_0 e^{2 i \kappa}\,(-3 + 3 i \kappa + \kappa^2)^2\,\lambda\,\Pi_0}{\kappa^4}.
\end{eqnarray}
These remarkably simple formulas, once plugged in Eq. \eqref{eq_ans5},
lead to a simple,
 exact expression for the parity violating vector mode function
 in the final epoch $\tau\ge\tau_2$ after the non-slow-roll phase occurs.

The amplitude of the resulting power
spectrum at the end of inflation, normalized by its value
at large scales, results
\begin{eqnarray}
{ \Pi}(\kappa)&=&\frac{{\cal P}_B(\kappa)}{{\cal P}_B(\kappa\to0)}
\\
&=&\big|C_1+C_2 \big|^2
\label{eq_sumC}
\end{eqnarray}
We plot it in Fig.~\ref{fig_simPB} for a representative choice of parameters. Notice that also in this analytical Scenario 2 we find evidence for
the phenomenon of `chirality flip' towards the peak, as found in Section~\ref{sec_scen1} for 
Scenario 1. More generally, for intermediate
values of the parameter $B_0$ connected
to the parity-violating effects, we notice that
the profiles of the resulting chiral spectra
resemble the numerical results of Section~\ref{sec_scen1}.
We conclude that the  analytical approach of this Section 
is able to capture the most important features of the system
in terms of very simple, analytical expressions for the parity-violating mode functions. 

\section{Chiral gravitational waves from magnetic fields}
\label{sec_vigw}

Our results so far indicate that  a short phase of non-slow-roll
dynamics can considerably enhance the spectrum of primordial  magnetic 
field at small scales. In this section we study the 
consequences of this phenomenon for
the production of a chiral stochastic gravitational
wave (GW) background  induced at the second order by the chiral magnetic field. 
After their production at the time of reentry during the epoch of radiation domination, the
gravitational waves propagate essentially freely due to the extreme weakness of their interactions, so that they traverse the perturbed cosmological background with negligible scattering or backreaction.
As a consequence, their subsequent evolution is largely decoupled from, and therefore insensitive to, the later dynamics and possible non-linear evolution of the magnetic fields, preserving to a very good approximation the information imprinted at the time of their generation.

The topic of vector induced GW has been developed
in various works \cite{Caprini:2001nb,Mack:2001gc}, and the role of helical magnetic
fields
investigated in \cite{Caprini:2014mja}. Our discussion aims to develop more systematic
arguments and general formulas that can be used in our  context
of non-slow-roll magnetogenesis. We generalize the approach of \cite{Atkins:2025pvg}
to study the generation of GW from the EM field during the radiation dominated epoch 
following inflation.
Recall that the conformal invariance of the EM field is restored post inflation,
since $J=1$ at the end of inflation.
The Fourier decomposition of the electromagnetic vector potential is
\begin{equation}
\label{eq_fou_new}
A_i(\tau,\mathbf{x})
=
\sum_{\lambda=L}^R
\int\!\frac{d^3\mathbf{k}}{(2\pi)^3}\,
e^{i\mathbf{k}\cdot\mathbf{x}}\,
e_i^{(\lambda)}(\hat{k})\,A_\lambda(\tau,\mathbf{k}),
\end{equation}
where $e_i^{(\lambda)}(\hat{k})$ are two orthonormal polarization vectors satisfying 
$e_i^{(\lambda)}\hat{k}_i=0$.  Vectors with a hat denote unit vectors, $\hat{k}=\mathbf{k}/k$ with $k\equiv|\mathbf{k}|$. 
The helicity polarization vectors obey the eigenrelation
\begin{equation}
\label{helicityeigen}
\big(i\varepsilon_{ijl}k_j\big)\,e_l^{(\lambda)}(\hat{k})=\xi^\lambda\,k\,e_i^{(\lambda)}(\hat{k}),
\qquad \xi^{\rm L}=-\xi^{\rm R}=1,
\end{equation}
and the normalization \(e_i^{(\lambda)}e_i^{(\sigma)*}=\delta^{\lambda\sigma}\).  
The comoving magnetic field is defined by
\begin{equation}
B_i(\tau,\mathbf{x})=\varepsilon_{ijk}\partial_j A_k(\tau,\mathbf{x}),
\qquad
B_i(\tau,\mathbf{k})=i\varepsilon_{ijl}k_j A_l(\tau,\mathbf{k}).
\end{equation}
Using \eqref{helicityeigen} and the expansion \eqref{eq_fou_new} we obtain
\begin{equation}
\label{eq:B_from_A_checked}
B_i(\tau,\mathbf{k})
= \sum_{\lambda=L,R} \xi^\lambda\,k\,e_i^{(\lambda)}(\hat{k})\,A_\lambda(\tau,\mathbf{k}).
\end{equation}
We assume the two helicities of the vector potential have (in general) different spectra:
\begin{equation}
\label{eq:PA_def_repeat}
\langle A_\lambda(\tau,\mathbf{k})\,A_\sigma^*(\tau,\mathbf{q})\rangle'
=\delta_{\lambda\sigma}\,\frac{2\pi^2}{k^3}\,\mathcal{P}_A^{\,\lambda}(k),
\end{equation}
where the prime over the expectation value denotes the quantity barring 
the associated Dirac delta function, here $\delta^{(3)}(\vk-\vq)$.
We define symmetric and helical combinations of ${\cal P}_A(k)$ as
\begin{equation}
\mathcal{P}_A^{S}(k)\equiv\frac{\mathcal{P}_A^{L}(k)+\mathcal{P}_A^{R}(k)}{2},
\qquad
\mathcal{P}_A^{H}(k)\equiv\frac{\mathcal{P}_A^{L}(k)-\mathcal{P}_A^{R}(k)}{2}.
\end{equation}
The quantity $\mathcal{P}_A^{H}(k)$ controls the amount of helicity in the vector sector. It is controlled by the quantity $\gamma$
in the specific scenarios discussed in Sections \ref{sec_scen1} and \ref{sec_scen2}. With these
ingredients
we can
compute the magnetic-field two-point  functions:
\begin{align}
\langle B_i(\mathbf{k})\,B_j^*(\mathbf{q})\rangle'
&=\sum_{\lambda,\sigma}\xi^\lambda\,\xi^\sigma\,k^2\,
e_i^{(\lambda)}(\hat{k})\,e_j^{(\sigma)\ast}(\hat{k})\,
\langle A_\lambda(\mathbf{k})A_\sigma^\ast(\mathbf{q})\rangle'
\nonumber\\
&=\sum_{\lambda}(\xi^\lambda)^2\,k^2\,
e_i^{(\lambda)}(\hat{k})\,e_j^{(\lambda)\ast}(\hat{k})\,
\frac{2\pi^2}{k^3}\,\mathcal{P}_A^{\,\lambda}(k)
\nonumber\\
&=\frac{2\pi^2}{k}\sum_{\lambda=L,R}\mathcal{P}_A^{\,\lambda}(k)\,
e_i^{(\lambda)}(\hat{k})\,e_j^{(\lambda)\ast}(\hat{k}).
\end{align}
We now make use of the helicity-basis identity
\begin{equation}
\label{eq:e_e_star_repeat}
e_i^{(\lambda)}(\hat{k})\,e_j^{(\lambda)*}(\hat{k})
=\frac{1}{2}\Big(\pi_{ij}(\hat{k})+i\xi^\lambda\,\varepsilon_{ijl}\hat{k}_l\Big),
\end{equation}
to obtain
\begin{align}
\langle B_i(\mathbf{k})\,B_j^*(\mathbf{q})\rangle'
&=\frac{2\pi^2}{k}\frac{1}{2}\Big[
\pi_{ij}(\hat{k})\big(\mathcal{P}_A^{L}(k)+\mathcal{P}_A^{R}(k)\big)
+i\,\varepsilon_{ijl}\hat{k}_l\big(\mathcal{P}_A^{L}(k)-\mathcal{P}_A^{R}(k)\big)
\Big]
\nonumber\\
&=\frac{2\pi^2}{k}\Big[
\pi_{ij}(\hat{k})\,\mathcal{P}_A^{S}(k)
+i\,\varepsilon_{ijl}\hat{k}_l\,\mathcal{P}_A^{H}(k)
\Big].
\label{eq:BB_from_PA_checked}
\end{align}
This  is the usual decomposition into parity-even (symmetric) and parity-odd parts: note the explicit factor \(i\) in front of the Levi-Civita tensor for the parity-odd contribution (this is required for the correlator to have the correct Hermiticity properties).

In order to map to the more common \(k^{-3}\) convention for the magnetic field spectrum,  we
define the associated magnetic spectra by
\begin{equation}
\label{eq:PB_mapping_checked}
{\qquad
\mathcal{P}_B^{S}(k)\equiv k^2\,\mathcal{P}_A^{S}(k), \qquad
\mathcal{P}_B^{H}(k)\equiv k^2\,\mathcal{P}_A^{H}(k).
\qquad}
\end{equation}
Then
\begin{equation}
\label{eq:BB_k3form_checked}
{\qquad
\langle B_i(\mathbf{k})\,B_j^*(\mathbf{q})\rangle'
=\frac{2\pi^2}{k^3}\left[
\pi_{ij}(\hat{k})\,\mathcal{P}_B^{S}(k)
+i\,\varepsilon_{ijl}\hat{k}_l\,\mathcal{P}_B^{H}(k)
\right].
\qquad}
\end{equation}

\smallskip
We now proceed computing the GW spectrum induced at second order by the enhanced magnetic field configuration.
After the end of inflation, tensor perturbations \(h_{ij}\) obey
\begin{equation}
\label{eq_strheq_new}
h_{ij}''(\tau,\mathbf{x})
+2{\cal H}(\tau)\,h_{ij}'(\tau,\mathbf{x})
-\nabla^2 h_{ij}(\tau,\mathbf{x})
= \Pi^{(T)}_{ij}(\tau,\mathbf{x}),
\end{equation}
where \(\Pi_{ij}^{(T)}\) is the transverse--traceless (TT) part of the anisotropic stress associated with  the amplified magnetic field.  
In Fourier space we decompose
\begin{equation}
h_{ij}(\tau,\mathbf{x})=
\sum_{\lambda=\pm2}\int\!\frac{d^3\mathbf{k}}{(2\pi)^3}\,
e^{i\mathbf{k}\cdot\mathbf{x}}\,
e^{(\lambda)}_{ij}(\mathbf{k})\,
h_{\mathbf{k}}^{(\lambda)}(\tau),
\end{equation}
where the spin-2 polarization tensors $e^{(\lambda)}_{ij}(\mathbf{k})$
with $\lambda=[L,R]=\pm2$, are given by the product of two spin-1 tensors used
to decompose the vector fields in Eq.~\eqref{eq_fou_new}:
$e^{(\lambda)}_{ij}=e^{(\lambda)}_{i} e^{(\lambda)}_{j}$.
The evolution equation for each Fourier mode for the spin-2 GW is
\begin{equation}
\label{eq_eveqh_new}
h_{\mathbf{k}}^{(\lambda)''}
+2{\cal H}h_{\mathbf{k}}^{(\lambda)'}
+k^2 h_{\mathbf{k}}^{(\lambda)}
= S^{(\lambda)}(\tau,\mathbf{k}),
\end{equation}
where the source term is the TT-projection of the magnetic stress:
\begin{eqnarray}
S^{(\lambda)}(\tau,\mathbf{k})
&\equiv& e^{(\lambda)\,ij}(\hat{k})\,\Pi^{(T)}_{ij}(\tau,\mathbf{k})
= \frac{e^{(\lambda)\,ij}(\hat{k})\,\Lambda_{ij}^{\,\,mn}\,
\tau_{mn}^{B}(\mathbf{k})}{a^2(\tau)}\,,
\nonumber
\\
&=& \frac{e^{(\lambda)\,ij}(\hat{k})\,
\tau_{ij}^{B}(\mathbf{k})}{a^2(\tau)}.
\label{eq_source_def}
\end{eqnarray}
To write these equalities 
 we used the definition of projector $\Lambda$:
\begin{equation}
\label{def_LAten_repeat}
\Lambda_{ij}^{\,\,\ell m}
=\frac12 \sum_{\lambda} e^{(\lambda)}_{ij}e^{(\lambda)\,\ell m}
 =\frac{1}{2}
\left(
\pi_i^\ell\pi_j^m+\pi_j^\ell\pi_i^m
-\pi_{ij}\pi^{\ell m}
\right),
\qquad
\pi_{ij}=\delta_{ij}-\hat{k}_i\hat{k}_j,
\end{equation}
The magnetic stress tensor is
\begin{equation}
\label{eq_emtmf}
\tau_{ij}^{(B)}({\bf k})
\,=\,\frac{1}{4 \pi}\int \frac{d^3 p}{(2\pi)^3}
\left[B_i({\bf p})B_j({\bf k}-{\bf p})-\frac{\delta_{ij}}{2}
B_m({\bf p})B_m({\bf k}-{\bf p})
\right].
\end{equation}
We ignore the contribution from electric field in the source as it gets
shorted during reheating, as mentioned earlier.
The formal solution of Eq.~\eqref{eq_eveqh_new} is
\begin{equation}
\label{eq_frh_new}
h_{\mathbf{k}}^{(\lambda)}(\tau)
=\frac{1}{a(\tau)}
\int d\tau'\,
g_k(\tau,\tau')\,a(\tau')\,S_{\mathbf{k}}^{(\lambda)}(\tau'),
\end{equation}
where \(g_k(\tau,\tau')\) is the Green function.  
During radiation domination the Green function is
\begin{equation}
g_k(\tau,\tau')=\frac{1}{k}\,
\big[\sin(k\tau)\cos(k\tau')-\sin(k\tau')\cos(k\tau)\big].
\end{equation}
The two point function (rescaled towards the standard power-spectrum definition) reads
\begin{align}
\frac{1}{2}\frac{k^3}{2\pi^2}
\langle
h_{\mathbf{k}}^{(\lambda)}(\tau)
h_{\mathbf{q}}^{(\sigma)*}(\tau)
\rangle'_{\mathbf{k}=\mathbf{q}}
&=
\frac{k^3}{4\pi^2 a^2(\tau)}
\int d\tau_1 d\tau_2\,
g_k(\tau,\tau_1)\,g_k(\tau,\tau_2)\,
a(\tau_1)a(\tau_2)
\nonumber\\
&\qquad\qquad
\times\;
\langle
S_{\mathbf{k}}^{(\lambda)}(\tau_1)
S_{\mathbf{k}}^{(\sigma)*}(\tau_2)
\rangle'.
\label{eq_ts2s_new}
\end{align}

We decompose GW correlators in terms of Stokes-like structures, focussing on intensity and circular polarization. Ignoring linear (spin-4) polarizations, the product of polarization tensors can be organized into parity-even and parity-odd parts as
\begin{equation}
\label{pro_pv}
e_{ij}^{(\lambda)} e_{mn}^{(\sigma)*}
= \frac{\delta^{\lambda\sigma}}{2}\,\Lambda_{ijmn}
+ i\frac{\varepsilon^{\lambda\sigma}}{2}\,\Sigma_{ijmn},
\end{equation}
where \(\Lambda_{ijmn}\) is the symmetric TT combination (given in \eqref{def_LAten_repeat}) 
and $\varepsilon^{\lambda\sigma}$ the Levi-Civita tensor in two dimensions with 
\(\varepsilon^{LR}=+1\). 
We conveniently define
\begin{equation}
X_{ij}(\hat v)\equiv\varepsilon_{ijl}\hat v_l,
\end{equation}
and express $\Sigma$ in the decomposition \eqref{pro_pv} as
\begin{equation}
\Sigma_{ijml}(\hat k)
=\frac12\big(\pi_{im}(\hat k)\,X_{jl}(\hat k)+\pi_{jl}(\hat k)\,X_{im}(\hat k)\big).
\end{equation}

\smallskip

We now collect these definitions and compute the source two-point function.
We have
\begin{eqnarray}
\big\langle S_{\mathbf{k}}^{(\lambda)}(\tau_1)\,
S_{\mathbf{k}}^{(\sigma)*}(\tau_2)\big\rangle'
&=&\frac{e_{ij}^{(\lambda)} e_{mn}^{(\sigma)*}}{a^2(\tau_1) a^2(\tau_2)}
\frac{(2 \pi^2)^2}{(4 \pi)^2}
\int \frac{d^3 p}{(2 \pi)^3\,p^3\,q^3}\,
\times
\nonumber
\\
&\times& \Big[ \left( \pi_{im}(\hat p) {\cal P}_B^S(p)+i X_{im}(\hat p)
{\cal P}_B^H(p)
\right) \left(
\pi_{jn}(\hat q) {\cal P}_B^S(q)+i X_{jn}(\hat q)
{\cal P}_B^H(q) \right)
\nonumber
\\
&+&\left( m\leftrightarrow q \right)
\Big]
\label{eq_geexS2}
\end{eqnarray}
with ${\bf q}={\bf k}-{\bf p}$. We decompose  $d^3 p = 2 \pi p^2 dp d \mu$, with $\mu = {\bf p}\cdot {\bf k}/(p k)$. We can write
\begin{align}
\label{eq:S_decomp}
\big\langle S_{\mathbf{k}}^{(\lambda)}(\tau_1)\,
S_{\mathbf{k}}^{(\sigma)*}(\tau_2)\big\rangle'
&= \frac{ k^{-3}}{4\,a^2(\tau_1)a^2(\tau_2)}\,
\Big[
\delta^{\lambda\sigma}\,\Gamma_{\rm even}(k)
-\varepsilon^{\lambda\sigma}\,\Gamma_{\rm odd}(k)
\Big],
\end{align}
where the scalar functions \(\Gamma_{\rm even}\) (parity-even) and \(\Gamma_{\rm odd}\) (parity-odd) are real. 
Taking into account of the  properties of the Levi-Civita tensor, they are
obtained by expanding Eq.~\eqref{eq_geexS2}, keeping the non-vanishing contributions:
\bea
\Gamma_{\rm even}(k)&=&\int \frac{d p \,d \mu}{p\,\, |{\bf q}|^3/k^3}
\Big\{
\Lambda_{ijmn}(\hat k) \left[\pi_{im}(p)
\pi_{jn}(q)+   \pi_{in}(p)
\pi_{jm}(q) \right]\, {\cal P}_B^S(p)\, {\cal P}_B^S(q)
\nonumber
\\
&&-\Lambda_{ijmn}(\hat k) \left[X_{im}(p)X_{jn}(q) + X_{in}(p)X_{jm}(q) \right]\, 
{\cal P}_B^H(p)\, {\cal P}_B^H(q)
\Big\}
\eea
and
\bea
\Gamma_{\rm odd}(k)&=&\int \frac{d p d \mu }{p \,\,|{\bf q}|^3/k^3}
\Big\{
\Sigma_{ijmn}(\hat k) \left[\pi_{im}(p) X_{jn}(q) + 
\pi_{in}(p) X_{jm}(q) \right]\, 
{\cal P}_B^S(p)\, {\cal P}_B^H(q)
\nonumber
\\
&&+\Lambda_{ijmn}(\hat k) \left[X_{im}(p) \pi_{jn}(q) + 
X_{in}(p) \pi_{jm}(q) \right]\, 
{\cal P}_B^H(p)\, {\cal P}_B^S(q)
\Big\}
\eea

\smallskip
Using the dimensionless variables
\begin{equation}
u=\frac{|\mathbf{k}-\mathbf{p}|}{k}, \qquad
v=\frac{p}{k}, \qquad
\mu=\frac{\mathbf{k}\cdot\mathbf{p}}{k\,p}
=\frac{1-u^2+v^2}{2v},
\end{equation}
we obtain the final convolution integrals
\begin{align}
\label{eq:A_def}
\Gamma_{\rm even}(k;\tau_1,\tau_2)
&= 
\int_0^\infty dv\,\int_{-1}^{1} d\mu\;\frac{1}{u^3 v}\;
\Big[\,\mathcal{C}_{SS}(u,v,\mu)\,\mathcal{P}_B^{S}(k u)\,\mathcal{P}_B^{S}(k v)\nonumber\\[-4pt]
&\qquad\qquad\qquad\qquad\qquad\qquad
+\;\mathcal{C}_{HH}(u,v,\mu)\,\mathcal{P}_B^{H}(k u)\,\mathcal{P}_B^{H}(k v)\,\Big],
\end{align}
and
\begin{align}
\label{eq:B_def}
\Gamma_{\rm odd}(k;\tau_1,\tau_2)
&=
\int_0^\infty dv\,\int_{-1}^{1} d\mu\;\frac{1}{u^3 v}\;
\;
\nonumber\\[-3pt]
&\times
\Big[\,\mathcal{C}^A_{\rm mix}(u,v,\mu)\,\mathcal{P}_B^{S}(k u)\,\mathcal{P}_B^{H}(k v)+
\mathcal{C}^B_{\rm mix}(u,v,\mu)\, \mathcal{P}_B^{H}(k u)\,\mathcal{P}_B^{S}(k v)\,\Big].
\end{align}
Here \(\mathcal{C}_{SS},\mathcal{C}_{HH},\mathcal{C}^{A,B}_{\rm mix}\) are scalar kernels obtained after contracting the geometric tensors with the TT projector and polarization tensors.
After performing the necessary contractions the scalar kernels result\\
\begin{equation}
\mathcal{C}_{SS}(u,v,\mu) = 
\frac{(1+\mu^{2})\left(2+v\,(v-4\mu+v\mu^{2})\right)}{2\left(1+v^{2}-2v\mu\right)}
\hskip0.5cm,\hskip0.5cm
\mathcal{C}_{HH}(u,v,\mu)
=\frac{2\mu\,(-1+v\mu)}{\sqrt{1+v^{2}-2v\mu}}
\label{eq:C_HH}
\end{equation}

\begin{equation}
\mathcal{C}^A_{\rm mix}(u,v,\mu)
=\frac{(1-v\mu)\,(1+\mu^{2})}{\sqrt{1+v^{2}-2v\mu}}
\hskip0.5cm, \hskip0.5cm
\mathcal{C}^B_{\rm mix}(u,v,\mu)
=\frac{\mu\left(2+v\,(v-4\mu+v\mu^{2})\right)}{1+v^{2}-2v\mu}
\label{eq:C_mixB}
\end{equation}
Plugging these formulas into \eqref{eq_ts2s_new}, the tensor two-point function can be compactly written as (a prime indicates correlators without the  Dirac
delta functions associated with momentum conservation)
\begin{equation}
\label{exp_fact}
\frac{1}{2}\frac{k^3}{2\pi^2}
\langle
h_{\mathbf{k}}^{(\lambda)}(\tau)
h_{\mathbf{s}}^{(\sigma)*}(\tau)
\rangle'_{\mathbf{k}=\mathbf{q}}
=\frac{\pi}{2\,a^2(\tau)}\,{\cal I}_\tau^2\,{\cal I}_{uv}^{(\lambda \sigma)},
\end{equation}
with
\begin{align}
{\cal I}_\tau^2&=
\left[
\int_{\tau_R}^{\tau}\!
\frac{d\tau_1}{a(\tau_1)}\,g_k(\tau,\tau_1)
\right]^2,
\label{def_IT_new}\\[3pt]
{\cal I}^{(\lambda \sigma)}_{uv}&=\delta^{\lambda\sigma}\,\Gamma_{\rm even}(k;\tau_1,\tau_2)
+\varepsilon^{\lambda\sigma}\,\Gamma_{\rm odd}(k;\tau_1,\tau_2).
\label{eq_iuv_new}
\end{align}
Notice that for vector-induced gravitational-wave spectra, the two-point correlator~\eqref{exp_fact} factorizes into a product of a time integral and a momentum integral: the former depends only on Green’s functions during radiation domination, while the latter encodes the properties of the magnetic-field source, see~\cite{Atkins:2025pvg}. This separation allows us to estimate the two contributions independently.

\paragraph{Time integral:}
During radiation domination, $a(\tau)/a(\tau_1)=\tau/\tau_1$ and $a(\tau)H(\tau)=1/\tau$.  
Averaging over rapid oscillations yields
\begin{equation}
\overline{{\cal I}_\tau^2}
=\frac{1}{2k^2 a^4 H^2}
\Big[
{\rm Ci}(-k\tau_R)^2
+\big(\tfrac{\pi}{2}-{\rm Si}(-k\tau_R)\big)^2
\Big],
\end{equation}
where ${\rm Ci}$ and ${\rm Si}$ are the cosine and sine integral functions. $\tau_R$ is a fiducial conformal 
time when the radiation domination epoch starts. 

\paragraph{Momentum integral:}
We introduce the standard variables $t$ and $s$ via
\begin{equation}
u=\frac{t+s+1}{2}, \qquad v=\frac{t-s+1}{2}.
\end{equation}
 Including the Jacobian for this transformation, we can carry out the momentum
integral for $0\le t \le \infty$ and $-1\le s \le 1$.\footnote{For numerical 
implementation, we take the range of $t$ to be $[10^{-2},200]$.}
The resulting GW energy density is (for the moment we keep
the polarization indexes, later we define more `physical' quantities)
\begin{equation}
\Omega_{\rm GW}^{(\lambda \sigma)}=
\frac{k^2}{12 a^2 H^2} \times \frac{1}{2}\frac{k^3}{2\pi^2}
\langle
h_{\mathbf{k}}^{(\lambda)}(\tau)
h_{\mathbf{q}}^{(\sigma)*}(\tau)
\rangle'
=
\frac{\pi k^2}{24\,a^4 H^2}\,
\overline{{\cal I}_\tau^2}\,
{\cal I}_{uv}^{(\lambda \sigma)}.
\label{eq_ts2sV2_new}
\end{equation}
During radiation domination,
\begin{equation}
a(\tau)=H_0\sqrt{\Omega_{\rm rd}}\,\tau,
\end{equation}
where $H_0$ and $\Omega_{\rm rd}$ are the present-day Hubble parameter and radiation energy fraction.  
Defining $\hat{\Omega}_B=\rho_B/\rho_{\rm cr}={\cal P}_B^{\rm CMB}/(3H_0^2)$ and including the redshift 
factor $\Omega_{\rm rd}$~\cite{Domenech:2021ztg}, we obtain
\begin{equation}
\label{eq_fiogw_new}
\Omega_{\rm GW}^{(\lambda \sigma)}(\kappa)
=
\left[\frac{3\,\hat{\Omega}_B^2}{64\,\Omega_{\rm rd}}\right]
\Big[
{\rm Ci}^2(\kappa x_\star)
+\big(\tfrac{\pi}{2}-{\rm Si}(\kappa x_\star)\big)^2
\Big]
\,{\cal I}^{(\lambda \sigma)}_{uv}(\kappa),
\end{equation}
with
 $\kappa=-k \tau_1$, with $|\tau_1|$ the epoch during inflation
 associated with  the non-slow-roll phase.  
The small parameter $x_\star=|\tau_R/\tau_1|$ controls the ratio between the relevant conformal times 
when radiation domination starts versus the non-slow-roll epoch,  and enters only logarithmically in 
the final result; we adopt $x_\star=10^{-4}$ for definiteness.
The prefactor $\left[{3\Omega_{\rm B}^2}/{64\Omega_{\rm rd}}\right]$ 
is fixed to be $2 \times 10^{-48}$ as per the arguments presented in~\cite{Atkins:2025pvg}.

We can now  introduce more physically-transparent combinations, by extracting the parity even and the parity odd  parts of the GW spectrum, defined as
\begin{eqnarray}
\Omega_{\rm GW}&=&\Omega_{\rm GW}^{(LL)}
+\Omega_{\rm GW}^{(RR)}
\\
\Omega_{\rm GW}^{ V}&=&\Omega_{\rm GW}^{(LL)}
-\Omega_{\rm GW}^{(RR)}
\end{eqnarray}
The quantity $\Omega_{\rm GW}$ is the traditional GW energy density computed
through the GW intensity, while $\Omega_{\rm GW}^{ V}$ aims at characterizing the effects of helical magnetic fields which induce
circular polarization in the GW background.

\begin{figure}[t!]
\includegraphics[width=0.48\linewidth]{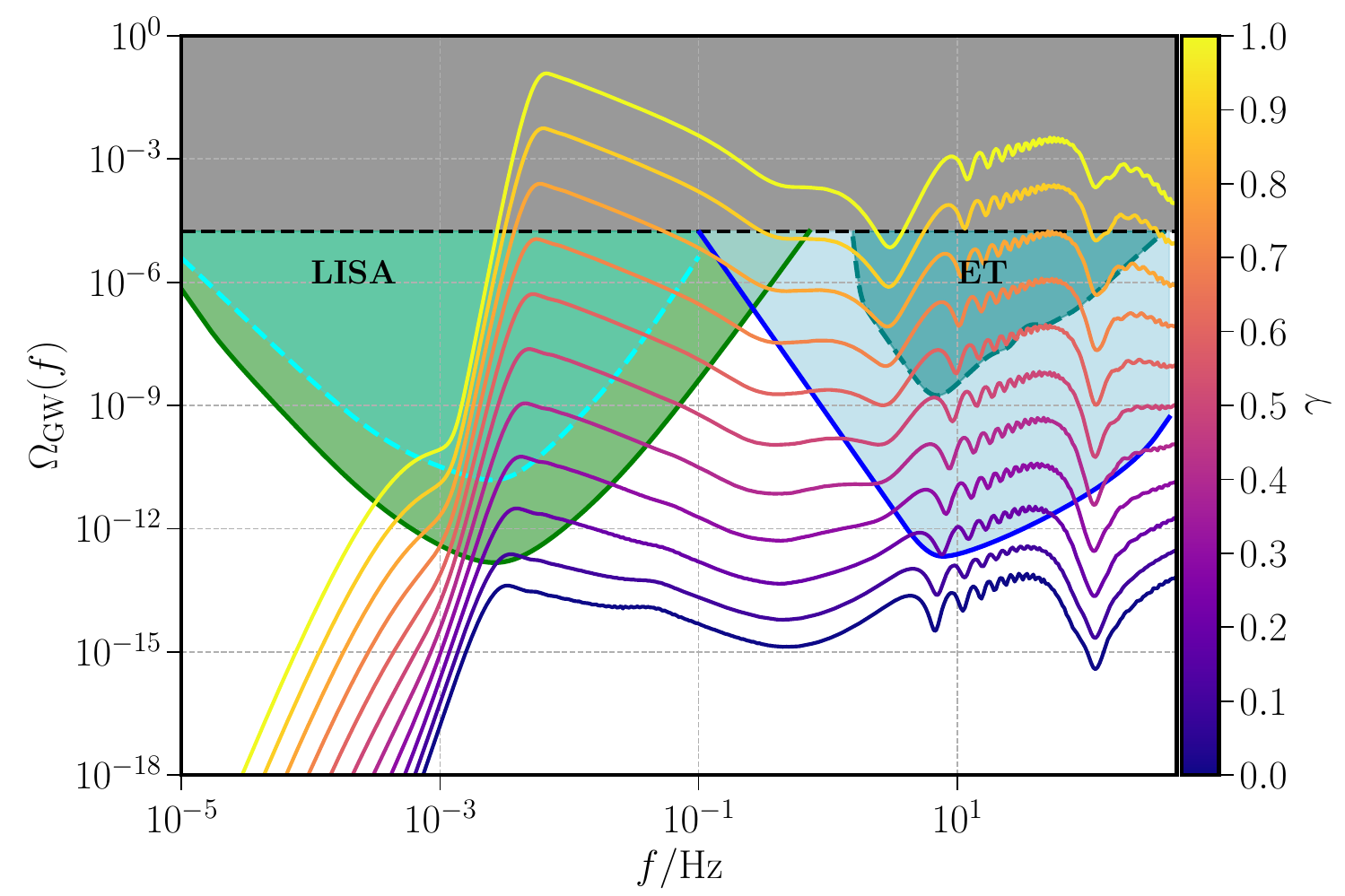}
\includegraphics[width=0.48\linewidth]{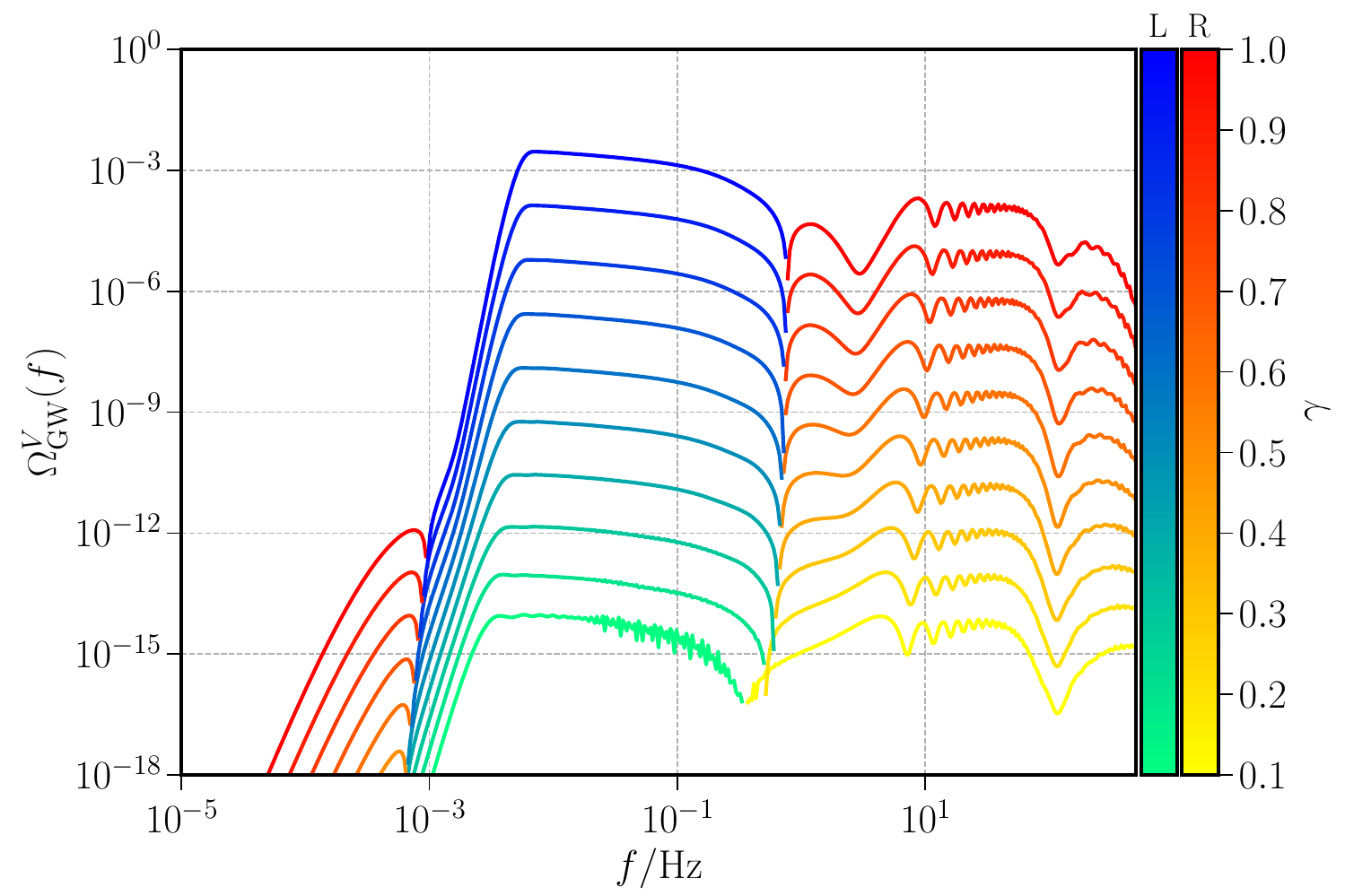}
\caption{\small The spectral density $\ogw$ and the associated V mode $\ogw^V$ are 
presented for a range of $\gamma=[0,1]$, corresponding to the magnetic spectra
presented in Fig.~\ref{fig:pb-vs-gamma}.
For reference, we plot the sensitivity curves of LISA (the nominal form 
in cyan and for the broken-power law case in green) and ET (the nominal form in teal 
and for the power-law case in blue)~\cite{Chowdhury:2022gdc,Marriott-Best:2024anh}.
We also demarcate the bound of $\ogw < 1.74 \times 10^{-5}$ from BBN (in black dashed 
line)~\cite{Smith:2006nka}.
Beyond the overall amplification of $\ogw$ due to non-zero $\gamma$, we find the 
spectrum having significant helicity as evidenced in $\ogw^V$.
The feature of altering helicity of $\pb(k)$ induces corresponding feature of
altering chirality around the peak of GW spectra as seen in $\ogw^V$.}
\label{fig:ogw-vgw-gamma}
\end{figure}

\subsection{Results and implications for GW experiments}

We  compute the GW density parameter $\ogw$ and associated chirality $\ogw^V$ induced by the spectra $\pb(k)$ using Eq.~\eqref{eq_fiogw_new}. We focus on our first scenario described in Section \ref{sec_scen1} -- see the plot 
 in Fig.~\ref{fig:pb-vs-gamma}.
The behavior of the resulting GW spectra are illustrated in 
Fig.~\ref{fig:ogw-vgw-gamma}, where we plot $\ogw$ and $\ogw^V$ as 
functions of frequency.
We  set $n_2=-3$, $\Delta N_2=3$ and the onset of the intermediate phase
in $J$ at
$N_1=15$ such that the corresponding wavenumber $k_1 \simeq 10^{14}\,\mpcinv$.
These  choices are motivated by the requirement that the peak of $\ogw$ enters in the 
 range of interest for future GW experiments, once the units are converted
into frequency expressed in  Hertz (more on this later).

\smallskip
A parity-violating term in the action, controlled by the parameter $\gamma$, significantly increases the 
induction of GW. See Fig.~\ref{fig:ogw-vgw-gamma}, left panel. 
For example, choosing
 $\gamma=1$ leads to  an amplification of about $10^{12}$ in the amplitude
 of $\ogw$,  
when compared to $\gamma=0$.
This behaviour  can be understood from the relation $\ogw \propto \pb^2$ and the amplification is about $10^6$ 
at the peak of $\pb$ [cf.~Fig.~\ref{fig:pb-vs-gamma}].
The bound from BBN already rules out a range of values of $\gamma>0.7$ for the case of
$n_2 = -3$ and $\Delta N_2=3$.
As to $\ogw(f)$, we find that the values of $\gamma \in [0.2,0.7]$ leads to amplitude 
significant enough to be detected by the sensitivity of LISA.
Besides, the features of the magnetic field spectral growth, its prominent peak, a second sub-dominant peak, and its 
subsequent oscillations as seen in $\pb(k)$ are well reflected and captured in the $\ogw(f)$ profile.

Turning to the behavior of chirality in the GW spectrum $\ogw^V(f)$, the signature of 
altering helicities around the peak of $\pb(k)$ is reproduced in the chirality of induced GW. See Fig.~\ref{fig:ogw-vgw-gamma}, right panel. 
Note that it is non-trivial to expect this behavior {\it a priori}, due to the non-linearities in the convolution integrals, and the specific integral  kernels.
We may see from the structure of the kernels involved in integration of source functions 
[cf.~Eqs.~\eqref{eq:A_def} and~\eqref{eq:B_def}] that both polarizations of $\pb^\lambda(k)$ 
contribute to both chiralities of $\ogw^\lambda(f)$.
Nonetheless, we find that the peak amplitude of $\ogw^V$ is positive (left-chiral) and
it switches to negative values (right-chiral) on either sides.
Moreover, we should note that for any given value of $\gamma$, the spectrum is never
maximally chiral, i.e. $\ogw^V/\ogw <1$.
The reason being again that both polarizations of $\pb^{\rm L,R}(k)$ contribute to a given 
chirality of induced GW.

\smallskip
Our findings open up interesting opportunities for GW experiments. It is well known that directly measuring GW circular polarization is challenging with planar detectors such as LISA \cite{LISACosmologyWorkingGroup:2022jok,LISA:2024hlh} and the Einstein Telescope (ET) \cite{ET:2025xjr}, see e.g.\cite{Smith:2016jqs} (as well as with pulsar timing arrays (see, e.g., \cite{Kato:2015bye})). Nevertheless,  the two panels in 
Fig.~\ref{fig:ogw-vgw-gamma} suggest at least two distinct strategies for detecting
effects of parity violation in the GW spectrum:
\begin{enumerate}
\item Information about chirality can be extracted from measurements of $\ogw$ with LISA and ET in synergy \cite{Marriott-Best:2024anh}. In particular, as shown in the left panel of Fig.~\ref{fig:ogw-vgw-gamma}, $\ogw$ exhibits a characteristic double-peak structure that would be accessible through a multiband detection with LISA and ET. Distinctive features of the second peak in the ET frequency band—such as the abrupt change in slope preceding the peak and the subsequent oscillatory behavior—depend sensitively on the value of the chirality parameter $\gamma$. Accurate measurements of the spectral shape of $\ogw$ could therefore allow one to infer $\gamma$, potentially extending the reconstruction techniques developed in \cite{LISACosmologyWorkingGroup:2025vdz,Ghaleb:2025xqn}.
\item  For suitable choices of parameters, the amplitude of $\ogw^V$ can be quite large, and at the same time $\ogw$ satisfies BBN bounds at frequencies
between the milli-Hertz and the tens of Hertz. Although $\ogw^V$ can not
be detected with measurements of the isotropic background, it can be measured
from  the properties of the kinematic dipole anisotropy
 \cite{Seto:2006hf,Seto:2006dz,Kato:2015bye,Domcke:2019zls,Cruz:2024esk}, which might be detectable in this regime.
\end{enumerate}
A detailed quantitative investigation of both these possibilities is left for future work.
Further, we must note that there are very few models of parity-violation that are known
to generate strong signal of $\ogw$ over a wide range of frequency with distinctly different 
chiralities of GW over different windows of 
frequencies (see~\cite{Ragavendra:2025svk,Dimastrogiovanni:2025snj} for such scenarios).
The narrow set of candidate models predicting altering chirality shall help in
breaking of degeneracy during comparison against observations.


\subsection*{Acknowledgments}
HVR acknowledges support by the MUR PRIN2022 Project ``BROWSEPOL: Beyond 
standaRd mOdel With coSmic microwavE background POLarization''-2022EJNZ53 financed by the European Union - Next Generation EU. 
HVR acknowledges the use of computing cluster under 
University of Padova Strategic Research Infrastructure Grant 2017:
``CAPRI: Calcolo ad Alte Prestazioni per la Ricerca e l’Innovazione''.
HVR thanks Ye Shifeng for valuable queries and comments.
GT is partially funded by the STFC
grants ST/T000813/1 and ST/X000648/1. 
LS wishes to thank the Indo-French Centre for the Promotion of 
Advanced Research (IFCPAR/CEFIPRA), New Delhi, India, for support of the proposal
6704-4 titled `Testing flavors of the early universe beyond vanilla models with 
cosmological observations’ under the Collaborative Scientific Research Programme.

\bigskip
{\small
\addcontentsline{toc}{section}{References}
\bibliographystyle{utphys}
\bibliography{refs_MagSlope}

\providecommand{\href}[2]{#2}\begingroup\raggedright\begin{thebibliography}{10}

\bibitem{Maleknejad:2012fw}
A.~Maleknejad, M.~M. Sheikh-Jabbari, and J.~Soda, ``{Gauge Fields and
  Inflation},'' \href{https://dx.doi.org/10.1016/j.physrep.2013.03.003}{{\em
  Phys. Rept.} {\bfseries 528} (2013) 161--261},
  \href{https://arxiv.org/abs/1212.2921}{{\ttfamily arXiv:1212.2921 [hep-th]}}.

\bibitem{Turner:1987bw}
M.~S. Turner and L.~M. Widrow, ``{Inflation Produced, Large Scale Magnetic
  Fields},'' \href{https://dx.doi.org/10.1103/PhysRevD.37.2743}{{\em Phys. Rev.
  D} {\bfseries 37} (1988) 2743}.

\bibitem{Ratra:1991bn}
B.~Ratra, ``{Cosmological 'seed' magnetic field from inflation},''
  \href{https://dx.doi.org/10.1086/186384}{{\em Astrophys. J. Lett.} {\bfseries
  391} (1992) L1--L4}.

\bibitem{Finelli:2000sh}
F.~Finelli and A.~Gruppuso, ``{Resonant amplification of gauge fields in
  expanding universe},''
  \href{https://dx.doi.org/10.1016/S0370-2693(01)00199-X}{{\em Phys. Lett. B}
  {\bfseries 502} (2001) 216--222},
  \href{https://arxiv.org/abs/hep-ph/0001231}{{\ttfamily
  arXiv:hep-ph/0001231}}.

\bibitem{Maroto:2000zu}
A.~L. Maroto, ``{Primordial magnetic fields from metric perturbations},''
  \href{https://dx.doi.org/10.1103/PhysRevD.64.083006}{{\em Phys. Rev. D}
  {\bfseries 64} (2001) 083006},
  \href{https://arxiv.org/abs/hep-ph/0008288}{{\ttfamily
  arXiv:hep-ph/0008288}}.

\bibitem{Maroto:2001ki}
A.~L. Maroto, ``{Cosmological magnetic fields induced by metric perturbations
  after inflation},'' in {\em {5th International Conference on Particle Physics
  and the Early Universe}}.
\newblock 11, 2001.
\newblock \href{https://arxiv.org/abs/hep-ph/0111268}{{\ttfamily
  arXiv:hep-ph/0111268}}.

\bibitem{Matarrese:2004kq}
S.~Matarrese, S.~Mollerach, A.~Notari, and A.~Riotto, ``{Large-scale magnetic
  fields from density perturbations},''
  \href{https://dx.doi.org/10.1103/PhysRevD.71.043502}{{\em Phys. Rev. D}
  {\bfseries 71} (2005) 043502},
  \href{https://arxiv.org/abs/astro-ph/0410687}{{\ttfamily
  arXiv:astro-ph/0410687}}.

\bibitem{Martin:2007ue}
J.~Martin and J.~Yokoyama, ``{Generation of Large-Scale Magnetic Fields in
  Single-Field Inflation},''
  \href{https://dx.doi.org/10.1088/1475-7516/2008/01/025}{{\em JCAP} {\bfseries
  01} (2008) 025}, \href{https://arxiv.org/abs/0711.4307}{{\ttfamily
  arXiv:0711.4307 [astro-ph]}}.

\bibitem{Demozzi:2009fu}
V.~Demozzi, V.~Mukhanov, and H.~Rubinstein, ``{Magnetic fields from
  inflation?},'' \href{https://dx.doi.org/10.1088/1475-7516/2009/08/025}{{\em
  JCAP} {\bfseries 08} (2009) 025},
  \href{https://arxiv.org/abs/0907.1030}{{\ttfamily arXiv:0907.1030
  [astro-ph.CO]}}.

\bibitem{Kanno:2009ei}
S.~Kanno, J.~Soda, and M.-a. Watanabe, ``{Cosmological Magnetic Fields from
  Inflation and Backreaction},''
  \href{https://dx.doi.org/10.1088/1475-7516/2009/12/009}{{\em JCAP} {\bfseries
  12} (2009) 009}, \href{https://arxiv.org/abs/0908.3509}{{\ttfamily
  arXiv:0908.3509 [astro-ph.CO]}}.

\bibitem{Bamba:2006ga}
K.~Bamba and M.~Sasaki, ``{Large-scale magnetic fields in the inflationary
  universe},'' \href{https://dx.doi.org/10.1088/1475-7516/2007/02/030}{{\em
  JCAP} {\bfseries 02} (2007) 030},
  \href{https://arxiv.org/abs/astro-ph/0611701}{{\ttfamily
  arXiv:astro-ph/0611701}}.

\bibitem{Bamba:2008hr}
K.~Bamba, C.~Q. Geng, and S.~H. Ho, ``{Large-scale magnetic fields from
  inflation due to Chern-Simons-like effective interaction},''
  \href{https://dx.doi.org/10.1088/1475-7516/2008/11/013}{{\em JCAP} {\bfseries
  11} (2008) 013}, \href{https://arxiv.org/abs/0806.1856}{{\ttfamily
  arXiv:0806.1856 [astro-ph]}}.

\bibitem{Barnaby:2012tk}
N.~Barnaby, R.~Namba, and M.~Peloso, ``{Observable non-gaussianity from gauge
  field production in slow roll inflation, and a challenging connection with
  magnetogenesis},'' \href{https://dx.doi.org/10.1103/PhysRevD.85.123523}{{\em
  Phys. Rev. D} {\bfseries 85} (2012) 123523},
  \href{https://arxiv.org/abs/1202.1469}{{\ttfamily arXiv:1202.1469
  [astro-ph.CO]}}.

\bibitem{Ferreira:2013sqa}
R.~J.~Z. Ferreira, R.~K. Jain, and M.~S. Sloth, ``{Inflationary magnetogenesis
  without the strong coupling problem},''
  \href{https://dx.doi.org/10.1088/1475-7516/2013/10/004}{{\em JCAP} {\bfseries
  10} (2013) 004}, \href{https://arxiv.org/abs/1305.7151}{{\ttfamily
  arXiv:1305.7151 [astro-ph.CO]}}.

\bibitem{Ferreira:2014hma}
R.~J.~Z. Ferreira, R.~K. Jain, and M.~S. Sloth, ``{Inflationary Magnetogenesis
  without the Strong Coupling Problem II: Constraints from CMB anisotropies and
  B-modes},'' \href{https://dx.doi.org/10.1088/1475-7516/2014/06/053}{{\em
  JCAP} {\bfseries 06} (2014) 053},
  \href{https://arxiv.org/abs/1403.5516}{{\ttfamily arXiv:1403.5516
  [astro-ph.CO]}}.

\bibitem{Graham:2015rva}
P.~W. Graham, J.~Mardon, and S.~Rajendran, ``{Vector Dark Matter from
  Inflationary Fluctuations},''
  \href{https://dx.doi.org/10.1103/PhysRevD.93.103520}{{\em Phys. Rev. D}
  {\bfseries 93} no.~10, (2016) 103520},
  \href{https://arxiv.org/abs/1504.02102}{{\ttfamily arXiv:1504.02102
  [hep-ph]}}.

\bibitem{Durrer:2013pga}
R.~Durrer and A.~Neronov, ``{Cosmological Magnetic Fields: Their Generation,
  Evolution and Observation},''
  \href{https://dx.doi.org/10.1007/s00159-013-0062-7}{{\em Astron. Astrophys.
  Rev.} {\bfseries 21} (2013) 62},
  \href{https://arxiv.org/abs/1303.7121}{{\ttfamily arXiv:1303.7121
  [astro-ph.CO]}}.

\bibitem{Subramanian:2015lua}
K.~Subramanian, ``{The origin, evolution and signatures of primordial magnetic
  fields},'' \href{https://dx.doi.org/10.1088/0034-4885/79/7/076901}{{\em Rept.
  Prog. Phys.} {\bfseries 79} no.~7, (2016) 076901},
  \href{https://arxiv.org/abs/1504.02311}{{\ttfamily arXiv:1504.02311
  [astro-ph.CO]}}.

\bibitem{Ozsoy:2023ryl}
O.~{\"O}zsoy and G.~Tasinato, ``{Inflation and Primordial Black Holes},''
  \href{https://dx.doi.org/10.3390/universe9050203}{{\em Universe} {\bfseries
  9} no.~5, (2023) 203}, \href{https://arxiv.org/abs/2301.03600}{{\ttfamily
  arXiv:2301.03600 [astro-ph.CO]}}.

\bibitem{Tripathy:2021sfb}
S.~Tripathy, D.~Chowdhury, R.~K. Jain, and L.~Sriramkumar, ``{Challenges in the
  choice of the nonconformal coupling function in inflationary
  magnetogenesis},'' \href{https://dx.doi.org/10.1103/PhysRevD.105.063519}{{\em
  Phys. Rev. D} {\bfseries 105} no.~6, (2022) 063519},
  \href{https://arxiv.org/abs/2111.01478}{{\ttfamily arXiv:2111.01478
  [astro-ph.CO]}}.

\bibitem{Tripathy:2022iev}
S.~Tripathy, D.~Chowdhury, H.~V. Ragavendra, R.~K. Jain, and L.~Sriramkumar,
  ``{Circumventing the challenges in the choice of the nonconformal coupling
  function in inflationary magnetogenesis},''
  \href{https://dx.doi.org/10.1103/PhysRevD.107.043501}{{\em Phys. Rev. D}
  {\bfseries 107} no.~4, (2023) 043501},
  \href{https://arxiv.org/abs/2211.05834}{{\ttfamily arXiv:2211.05834
  [astro-ph.CO]}}.

\bibitem{Atkins:2025pvg}
B.~Atkins, D.~Chowdhury, A.~Marriott-Best, and G.~Tasinato, ``{Inflationary
  magnetogenesis beyond slow-roll and its induced gravitational waves},''
  \href{https://arxiv.org/abs/2507.01772}{{\ttfamily arXiv:2507.01772
  [astro-ph.CO]}}.

\bibitem{Tasinato:2020vdk}
G.~Tasinato, ``{An analytic approach to non-slow-roll inflation},''
  \href{https://dx.doi.org/10.1103/PhysRevD.103.023535}{{\em Phys. Rev. D}
  {\bfseries 103} no.~2, (2021) 023535},
  \href{https://arxiv.org/abs/2012.02518}{{\ttfamily arXiv:2012.02518
  [hep-th]}}.

\bibitem{Tasinato:2023ukp}
G.~Tasinato, ``{Large |{\ensuremath{\eta}}| approach to single field
  inflation},'' \href{https://dx.doi.org/10.1103/PhysRevD.108.043526}{{\em
  Phys. Rev. D} {\bfseries 108} no.~4, (2023) 043526},
  \href{https://arxiv.org/abs/2305.11568}{{\ttfamily arXiv:2305.11568
  [hep-th]}}.

\bibitem{Sorbo:2011rzq}
L.~Sorbo, ``{Parity violation in the Cosmic Microwave Background from a
  pseudoscalar inflaton},''
  \href{https://dx.doi.org/10.1088/1475-7516/2011/06/003}{{\em JCAP} {\bfseries
  06} (2011) 003}, \href{https://arxiv.org/abs/1101.1525}{{\ttfamily
  arXiv:1101.1525 [astro-ph.CO]}}.

\bibitem{Caprini:2014mja}
C.~Caprini and L.~Sorbo, ``{Adding helicity to inflationary magnetogenesis},''
  \href{https://dx.doi.org/10.1088/1475-7516/2014/10/056}{{\em JCAP} {\bfseries
  10} (2014) 056}, \href{https://arxiv.org/abs/1407.2809}{{\ttfamily
  arXiv:1407.2809 [astro-ph.CO]}}.

\bibitem{Ng:2014lyb}
K.-W. Ng, S.-L. Cheng, and W.~Lee, ``{Inflationary dilaton-axion
  magnetogenesis},'' \href{https://dx.doi.org/10.6122/CJP.20150909}{{\em Chin.
  J. Phys.} {\bfseries 53} (2015) 110105},
  \href{https://arxiv.org/abs/1409.2656}{{\ttfamily arXiv:1409.2656
  [astro-ph.CO]}}.

\bibitem{Tasinato:2014fia}
G.~Tasinato, ``{A scenario for inflationary magnetogenesis without strong
  coupling problem},''
  \href{https://dx.doi.org/10.1088/1475-7516/2015/03/040}{{\em JCAP} {\bfseries
  03} (2015) 040}, \href{https://arxiv.org/abs/1411.2803}{{\ttfamily
  arXiv:1411.2803 [hep-th]}}.

\bibitem{Subramanian:2009fu}
K.~Subramanian, ``{Magnetic fields in the early universe},''
  \href{https://dx.doi.org/10.1002/asna.200911312}{{\em Astron. Nachr.}
  {\bfseries 331} (2010) 110--120},
  \href{https://arxiv.org/abs/0911.4771}{{\ttfamily arXiv:0911.4771
  [astro-ph.CO]}}.

\bibitem{Sharma:2018kgs}
R.~Sharma, K.~Subramanian, and T.~R. Seshadri, ``{Generation of helical
  magnetic field in a viable scenario of inflationary magnetogenesis},''
  \href{https://dx.doi.org/10.1103/PhysRevD.97.083503}{{\em Phys. Rev. D}
  {\bfseries 97} no.~8, (2018) 083503},
  \href{https://arxiv.org/abs/1802.04847}{{\ttfamily arXiv:1802.04847
  [astro-ph.CO]}}.

\bibitem{Chowdhury:2018mhj}
D.~Chowdhury, L.~Sriramkumar, and M.~Kamionkowski, ``{Enhancing the
  cross-correlations between magnetic fields and scalar perturbations through
  parity violation},''
  \href{https://dx.doi.org/10.1088/1475-7516/2018/10/031}{{\em JCAP} {\bfseries
  10} (2018) 031}, \href{https://arxiv.org/abs/1807.07477}{{\ttfamily
  arXiv:1807.07477 [astro-ph.CO]}}.

\bibitem{Okano:2020uyr}
S.~Okano and T.~Fujita, ``{Chiral Gravitational Waves Produced in a Helical
  Magnetogenesis Model},''
  \href{https://dx.doi.org/10.1088/1475-7516/2021/03/026}{{\em JCAP} {\bfseries
  03} (2021) 026}, \href{https://arxiv.org/abs/2005.13833}{{\ttfamily
  arXiv:2005.13833 [astro-ph.CO]}}.

\bibitem{Brandenburg:2021bfx}
A.~Brandenburg, Y.~He, and R.~Sharma, ``{Simulations of Helical Inflationary
  Magnetogenesis and Gravitational Waves},''
  \href{https://dx.doi.org/10.3847/1538-4357/ac20d9}{{\em Astrophys. J.}
  {\bfseries 922} no.~2, (2021) 192},
  \href{https://arxiv.org/abs/2107.12333}{{\ttfamily arXiv:2107.12333
  [astro-ph.CO]}}.

\bibitem{Tripathy:2024ngu}
S.~Tripathy, D.~Chowdhury, H.~V. Ragavendra, and L.~Sriramkumar,
  ``{Cross-correlation between the curvature perturbations and magnetic fields
  in pure ultraslow roll inflation},''
  \href{https://dx.doi.org/10.1103/PhysRevD.111.063550}{{\em Phys. Rev. D}
  {\bfseries 111} no.~6, (2025) 063550},
  \href{https://arxiv.org/abs/2409.20232}{{\ttfamily arXiv:2409.20232
  [astro-ph.CO]}}.

\bibitem{Papanikolaou:2024cwr}
T.~Papanikolaou, C.~Tzerefos, S.~Capozziello, and G.~Lambiase,
  ``{Gravitational-wave signatures of gravito-electromagnetic couplings},''
  \href{https://dx.doi.org/10.1088/1475-7516/2025/01/051}{{\em JCAP} {\bfseries
  01} (2025) 051}, \href{https://arxiv.org/abs/2408.17259}{{\ttfamily
  arXiv:2408.17259 [astro-ph.CO]}}.

\bibitem{Ragavendra:2023ret}
H.~V. Ragavendra and L.~Sriramkumar, ``{Observational Imprints of Enhanced
  Scalar Power on Small Scales in Ultra Slow Roll Inflation and Associated
  Non-Gaussianities},'' \href{https://dx.doi.org/10.3390/galaxies11010034}{{\em
  Galaxies} {\bfseries 11} no.~1, (2023) 34},
  \href{https://arxiv.org/abs/2301.08887}{{\ttfamily arXiv:2301.08887
  [astro-ph.CO]}}.

\bibitem{Bhaumik:2019tvl}
N.~Bhaumik and R.~K. Jain, ``{Primordial black holes dark matter from
  inflection point models of inflation and the effects of reheating},''
  \href{https://arxiv.org/abs/1907.04125}{{\ttfamily arXiv:1907.04125
  [astro-ph.CO]}}.
[JCAP2001,037(2020)].

\bibitem{Ragavendra:2020sop}
H.~V. Ragavendra, P.~Saha, L.~Sriramkumar, and J.~Silk, ``{Primordial black
  holes and secondary gravitational waves from ultraslow roll and punctuated
  inflation},'' \href{https://dx.doi.org/10.1103/PhysRevD.103.083510}{{\em
  Phys. Rev. D} {\bfseries 103} no.~8, (2021) 083510},
  \href{https://arxiv.org/abs/2008.12202}{{\ttfamily arXiv:2008.12202
  [astro-ph.CO]}}.

\bibitem{Marriott-Best:2025sez}
A.~Marriott-Best, M.~Peloso, and G.~Tasinato, ``{New gravitational wave probe
  of vector dark matter},''
  \href{https://dx.doi.org/10.1103/PhysRevD.111.103511}{{\em Phys. Rev. D}
  {\bfseries 111} no.~10, (2025) 103511},
  \href{https://arxiv.org/abs/2502.13116}{{\ttfamily arXiv:2502.13116
  [astro-ph.CO]}}.

\bibitem{LaRosa:2025woi}
M.~La~Rosa and G.~Tasinato, ``{Ultralight dark matter from non-slow-roll
  inflation},'' \href{https://dx.doi.org/10.1103/p5nw-7k6z}{{\em Phys. Rev. D}
  {\bfseries 112} no.~12, (2025) 123542},
  \href{https://arxiv.org/abs/2508.16455}{{\ttfamily arXiv:2508.16455
  [astro-ph.CO]}}.

\bibitem{Byrnes:2018txb}
C.~T. Byrnes, P.~S. Cole, and S.~P. Patil, ``{Steepest growth of the power
  spectrum and primordial black holes},''
  \href{https://dx.doi.org/10.1088/1475-7516/2019/06/028}{{\em JCAP} {\bfseries
  06} (2019) 028}, \href{https://arxiv.org/abs/1811.11158}{{\ttfamily
  arXiv:1811.11158 [astro-ph.CO]}}.

\bibitem{Ozsoy:2019lyy}
O.~{\"O}zsoy and G.~Tasinato, ``{On the slope of the curvature power spectrum
  in non-attractor inflation},''
  \href{https://dx.doi.org/10.1088/1475-7516/2020/04/048}{{\em JCAP} {\bfseries
  04} (2020) 048}, \href{https://arxiv.org/abs/1912.01061}{{\ttfamily
  arXiv:1912.01061 [astro-ph.CO]}}.

\bibitem{Ragavendra:2024yfp}
H.~V. Ragavendra, A.~K. Sarkar, and S.~K. Sethi, ``{Constraining ultra slow
  roll inflation using cosmological datasets},''
  \href{https://dx.doi.org/10.1088/1475-7516/2024/07/088}{{\em JCAP} {\bfseries
  07} (2024) 088}, \href{https://arxiv.org/abs/2404.00933}{{\ttfamily
  arXiv:2404.00933 [astro-ph.CO]}}.

\bibitem{Talebian:2025jeg}
A.~Talebian and H.~Firouzjahi, ``{Axion ultraslow-roll inflation},''
  \href{https://dx.doi.org/10.1103/xxqr-b2cw}{{\em Phys. Rev. D} {\bfseries
  112} no.~12, (2025) 123548},
  \href{https://arxiv.org/abs/2507.02685}{{\ttfamily arXiv:2507.02685
  [astro-ph.CO]}}.

\bibitem{MAGIC:2022piy}
{\bfseries MAGIC} Collaboration, V.~A. Acciari {\em et~al.}, ``{A lower bound
  on intergalactic magnetic fields from time variability of 1ES 0229+200 from
  MAGIC and Fermi/LAT observations},''
  \href{https://dx.doi.org/10.1051/0004-6361/202244126}{{\em Astron.
  Astrophys.} {\bfseries 670} (2023) A145},
  \href{https://arxiv.org/abs/2210.03321}{{\ttfamily arXiv:2210.03321
  [astro-ph.HE]}}.

\bibitem{Burmeister:2025lgo}
L.~Burmeister, P.~Da~Vela, F.~Longo, G.~Marti-Devesa, M.~Meyer, F.~Saturni,
  A.~Stamerra, and P.~Veres, ``{Constraints on the intergalactic magnetic field
  from Fermi-LAT observations of GRB 221009A},''
  \href{https://arxiv.org/abs/2512.11128}{{\ttfamily arXiv:2512.11128
  [astro-ph.HE]}}.

\bibitem{Zucca:2016iur}
A.~Zucca, Y.~Li, and L.~Pogosian, ``{Constraints on Primordial Magnetic Fields
  from Planck combined with the South Pole Telescope CMB B-mode polarization
  measurements},'' \href{https://dx.doi.org/10.1103/PhysRevD.95.063506}{{\em
  Phys. Rev. D} {\bfseries 95} no.~6, (2017) 063506},
  \href{https://arxiv.org/abs/1611.00757}{{\ttfamily arXiv:1611.00757
  [astro-ph.CO]}}.

\bibitem{Paoletti:2018uic}
D.~Paoletti, J.~Chluba, F.~Finelli, and J.~A. Rubino-Martin, ``{Improved CMB
  anisotropy constraints on primordial magnetic fields from the
  post-recombination ionization history},''
  \href{https://dx.doi.org/10.1093/mnras/sty3521}{{\em Mon. Not. Roy. Astron.
  Soc.} {\bfseries 484} no.~1, (2019) 185--195},
  \href{https://arxiv.org/abs/1806.06830}{{\ttfamily arXiv:1806.06830
  [astro-ph.CO]}}.

\bibitem{Paoletti:2019pdi}
D.~Paoletti and F.~Finelli, ``{Constraints on primordial magnetic fields from
  magnetically-induced perturbations: current status and future perspectives
  with LiteBIRD and future ground based experiments},''
  \href{https://dx.doi.org/10.1088/1475-7516/2019/11/028}{{\em JCAP} {\bfseries
  11} (2019) 028}, \href{https://arxiv.org/abs/1910.07456}{{\ttfamily
  arXiv:1910.07456 [astro-ph.CO]}}.

\bibitem{Durrer:2010mq}
R.~Durrer, L.~Hollenstein, and R.~K. Jain, ``{Can slow roll inflation induce
  relevant helical magnetic fields?},''
  \href{https://dx.doi.org/10.1088/1475-7516/2011/03/037}{{\em JCAP} {\bfseries
  03} (2011) 037}, \href{https://arxiv.org/abs/1005.5322}{{\ttfamily
  arXiv:1005.5322 [astro-ph.CO]}}.

\bibitem{Urban:2011bu}
F.~R. Urban, ``{On inflating magnetic fields, and the backreactions thereof},''
  \href{https://dx.doi.org/10.1088/1475-7516/2011/12/012}{{\em JCAP} {\bfseries
  12} (2011) 012}, \href{https://arxiv.org/abs/1111.1006}{{\ttfamily
  arXiv:1111.1006 [astro-ph.CO]}}.

\bibitem{Bartolo:2015dga}
N.~Bartolo, S.~Matarrese, M.~Peloso, and M.~Shiraishi, ``{Parity-violating CMB
  correlators with non-decaying statistical anisotropy},''
  \href{https://dx.doi.org/10.1088/1475-7516/2015/07/039}{{\em JCAP} {\bfseries
  07} (2015) 039}, \href{https://arxiv.org/abs/1505.02193}{{\ttfamily
  arXiv:1505.02193 [astro-ph.CO]}}.

\bibitem{tHooft:1973alw}
G.~'t~Hooft, ``{A Planar Diagram Theory for Strong Interactions},''
  \href{https://dx.doi.org/10.1016/0550-3213(74)90154-0}{{\em Nucl. Phys. B}
  {\bfseries 72} (1974) 461}.

\bibitem{Caprini:2001nb}
C.~Caprini and R.~Durrer, ``{Gravitational wave production: A Strong constraint
  on primordial magnetic fields},''
  \href{https://dx.doi.org/10.1103/PhysRevD.65.023517}{{\em Phys. Rev. D}
  {\bfseries 65} (2001) 023517},
  \href{https://arxiv.org/abs/astro-ph/0106244}{{\ttfamily
  arXiv:astro-ph/0106244}}.

\bibitem{Mack:2001gc}
A.~Mack, T.~Kahniashvili, and A.~Kosowsky, ``{Microwave background signatures
  of a primordial stochastic magnetic field},''
  \href{https://dx.doi.org/10.1103/PhysRevD.65.123004}{{\em Phys. Rev. D}
  {\bfseries 65} (2002) 123004},
  \href{https://arxiv.org/abs/astro-ph/0105504}{{\ttfamily
  arXiv:astro-ph/0105504}}.

\bibitem{Domenech:2021ztg}
G.~Dom{\`e}nech, ``{Scalar Induced Gravitational Waves Review},''
  \href{https://dx.doi.org/10.3390/universe7110398}{{\em Universe} {\bfseries
  7} no.~11, (2021) 398}, \href{https://arxiv.org/abs/2109.01398}{{\ttfamily
  arXiv:2109.01398 [gr-qc]}}.

\bibitem{Chowdhury:2022gdc}
D.~Chowdhury, G.~Tasinato, and I.~Zavala, ``{The rise of the primordial tensor
  spectrum from an early scalar-tensor epoch},''
  \href{https://dx.doi.org/10.1088/1475-7516/2022/08/010}{{\em JCAP} {\bfseries
  08} no.~08, (2022) 010}, \href{https://arxiv.org/abs/2204.10218}{{\ttfamily
  arXiv:2204.10218 [gr-qc]}}.

\bibitem{Marriott-Best:2024anh}
A.~Marriott-Best, D.~Chowdhury, A.~Ghoshal, and G.~Tasinato, ``{Exploring
  cosmological gravitational wave backgrounds through the synergy of LISA and
  the Einstein Telescope},''
  \href{https://dx.doi.org/10.1103/PhysRevD.111.103001}{{\em Phys. Rev. D}
  {\bfseries 111} no.~10, (2025) 103001},
  \href{https://arxiv.org/abs/2409.02886}{{\ttfamily arXiv:2409.02886
  [astro-ph.CO]}}.

\bibitem{Smith:2006nka}
T.~L. Smith, E.~Pierpaoli, and M.~Kamionkowski, ``{A new cosmic microwave
  background constraint to primordial gravitational waves},''
  \href{https://dx.doi.org/10.1103/PhysRevLett.97.021301}{{\em Phys. Rev.
  Lett.} {\bfseries 97} (2006) 021301},
  \href{https://arxiv.org/abs/astro-ph/0603144}{{\ttfamily
  arXiv:astro-ph/0603144}}.

\bibitem{LISACosmologyWorkingGroup:2022jok}
{\bfseries LISA Cosmology Working Group} Collaboration, P.~Auclair {\em
  et~al.}, ``{Cosmology with the Laser Interferometer Space Antenna},''
  \href{https://dx.doi.org/10.1007/s41114-023-00045-2}{{\em Living Rev. Rel.}
  {\bfseries 26} no.~1, (2023) 5},
  \href{https://arxiv.org/abs/2204.05434}{{\ttfamily arXiv:2204.05434
  [astro-ph.CO]}}.

\bibitem{LISA:2024hlh}
{\bfseries LISA} Collaboration, M.~Colpi {\em et~al.}, ``{LISA Definition Study
  Report},'' \href{https://arxiv.org/abs/2402.07571}{{\ttfamily
  arXiv:2402.07571 [astro-ph.CO]}}.

\bibitem{ET:2025xjr}
{\bfseries ET} Collaboration, A.~Abac {\em et~al.}, ``{The Science of the
  Einstein Telescope},'' \href{https://arxiv.org/abs/2503.12263}{{\ttfamily
  arXiv:2503.12263 [gr-qc]}}.

\bibitem{Smith:2016jqs}
T.~L. Smith and R.~Caldwell, ``{Sensitivity to a Frequency-Dependent Circular
  Polarization in an Isotropic Stochastic Gravitational Wave Background},''
  \href{https://dx.doi.org/10.1103/PhysRevD.95.044036}{{\em Phys. Rev. D}
  {\bfseries 95} no.~4, (2017) 044036},
  \href{https://arxiv.org/abs/1609.05901}{{\ttfamily arXiv:1609.05901
  [gr-qc]}}.

\bibitem{Kato:2015bye}
R.~Kato and J.~Soda, ``{Probing circular polarization in stochastic
  gravitational wave background with pulsar timing arrays},''
  \href{https://dx.doi.org/10.1103/PhysRevD.93.062003}{{\em Phys. Rev. D}
  {\bfseries 93} no.~6, (2016) 062003},
  \href{https://arxiv.org/abs/1512.09139}{{\ttfamily arXiv:1512.09139
  [gr-qc]}}.

\bibitem{LISACosmologyWorkingGroup:2025vdz}
{\bfseries LISA Cosmology Working Group} Collaboration, J.~E. Gammal {\em
  et~al.}, ``{Reconstructing primordial curvature perturbations via
  scalar-induced gravitational waves with LISA},''
  \href{https://dx.doi.org/10.1088/1475-7516/2025/05/062}{{\em JCAP} {\bfseries
  05} (2025) 062}, \href{https://arxiv.org/abs/2501.11320}{{\ttfamily
  arXiv:2501.11320 [astro-ph.CO]}}.

\bibitem{Ghaleb:2025xqn}
A.~Ghaleb, A.~Malhotra, G.~Tasinato, and I.~Zavala, ``{Bayesian reconstruction
  of primordial perturbations from induced gravitational waves},''
  \href{https://dx.doi.org/10.1103/n23j-5bfc}{{\em Phys. Rev. D} {\bfseries
  112} no.~12, (2025) 123538},
  \href{https://arxiv.org/abs/2505.22534}{{\ttfamily arXiv:2505.22534
  [astro-ph.CO]}}.

\bibitem{Seto:2006hf}
N.~Seto, ``{Prospects for direct detection of circular polarization of
  gravitational-wave background},''
  \href{https://dx.doi.org/10.1103/PhysRevLett.97.151101}{{\em Phys. Rev.
  Lett.} {\bfseries 97} (2006) 151101},
  \href{https://arxiv.org/abs/astro-ph/0609504}{{\ttfamily
  arXiv:astro-ph/0609504}}.

\bibitem{Seto:2006dz}
N.~Seto, ``{Quest for circular polarization of gravitational wave background
  and orbits of laser interferometers in space},''
  \href{https://dx.doi.org/10.1103/PhysRevD.75.061302}{{\em Phys. Rev. D}
  {\bfseries 75} (2007) 061302},
  \href{https://arxiv.org/abs/astro-ph/0609633}{{\ttfamily
  arXiv:astro-ph/0609633}}.

\bibitem{Domcke:2019zls}
V.~Domcke, J.~Garcia-Bellido, M.~Peloso, M.~Pieroni, A.~Ricciardone, L.~Sorbo,
  and G.~Tasinato, ``{Measuring the net circular polarization of the stochastic
  gravitational wave background with interferometers},''
  \href{https://dx.doi.org/10.1088/1475-7516/2020/05/028}{{\em JCAP} {\bfseries
  05} (2020) 028}, \href{https://arxiv.org/abs/1910.08052}{{\ttfamily
  arXiv:1910.08052 [astro-ph.CO]}}.

\bibitem{Cruz:2024esk}
N.~M.~J. Cruz, A.~Malhotra, G.~Tasinato, and I.~Zavala, ``{Measuring the
  circular polarization of gravitational waves with pulsar timing arrays},''
  \href{https://dx.doi.org/10.1103/PhysRevD.110.103505}{{\em Phys. Rev. D}
  {\bfseries 110} no.~10, (2024) 103505},
  \href{https://arxiv.org/abs/2406.04957}{{\ttfamily arXiv:2406.04957
  [astro-ph.CO]}}.

\bibitem{Ragavendra:2025svk}
H.~V. Ragavendra and N.~Bartolo, ``{Twisted echoes of an odd quartet:
  Scalar-induced gravitational waves as a probe of primordial
  parity-violation},'' \href{https://arxiv.org/abs/2507.02733}{{\ttfamily
  arXiv:2507.02733 [astro-ph.CO]}}.

\bibitem{Dimastrogiovanni:2025snj}
E.~Dimastrogiovanni, M.~Fasiello, A.~Papageorgiou, and C.~Z. Gatica, ``{Pure
  chromo-natural inflation: signatures of particle production from weak to
  strong backreaction},''
  \href{https://dx.doi.org/10.1088/1475-7516/2025/09/042}{{\em JCAP} {\bfseries
  09} (2025) 042}, \href{https://arxiv.org/abs/2504.17750}{{\ttfamily
  arXiv:2504.17750 [astro-ph.CO]}}.

\end{thebibliography}\endgroup
}
\end{document}